\documentclass[12pt]{iopart}

\usepackage{iopams}

\makeatletter
\@namedef{ver@amsmath.sty}{}
\makeatother

\usepackage{siunitx}
\usepackage{isotope}
\usepackage{fixltx2e}
\usepackage{booktabs}
\usepackage{xcolor}
\usepackage{multirow}
\usepackage{url}
\usepackage{graphicx}
\usepackage{tikz}
\usepackage{tabularx}
\usepackage[all=normal,bibbreaks=tight,floats=tight,mathspacing=tight,wordspacing=tight]{savetrees}

\providecommand{\keywords}[1]{\textbf{\textit{Index terms---}} #1}
\newcommand{\DDO}{D\textsubscript{2}O}
\definecolor{datacolor}{RGB}{72,36,117}

\graphicspath{{./fig_final_report/}{./fig/}}

\newcommand{\surrogateplot}[6]{
\begin{tikzpicture}
	\node[anchor=south west,inner sep=0] at (0,0){\includegraphics[width=\textwidth]{#1}};
	\draw[black,fill=white,line
		width=0.05mm,align=right,font=\fontsize{5.5}{6.5}\selectfont]
		(0.42,#6) rectangle
		(2.55,2.7) node[pos=.5] {
			\\[0.1ex]
			$\mathrm{MAE} = \num{#2}$\\
			$S = \num{#3}$\\
			$R^2 = \num{#4}$\\
			$R^2_{\text{adj.}} = \num{#5}$
		};
	\draw[black,fill=white,line
		width=0.05mm,align=left,font=\fontsize{5.5}{7.5}\selectfont]
		(2.02,1.3) rectangle
		(4.3,0.38) node[pos=.5] {
			{\color{red}
			\leavevmode\leaders\hrule height 0.7ex
			depth\dimexpr0.4pt-0.7ex\hskip3pt\kern0pt
			\hskip1pt
			\leavevmode\leaders\hrule height 0.7ex
			depth\dimexpr0.4pt-0.7ex\hskip3pt\kern0pt
			} Ideal model\\
			{\leavevmode\leaders\hrule height 0.7ex
			depth\dimexpr0.4pt-0.7ex\hskip3pt\kern0pt
			\hskip1pt
			\leavevmode\leaders\hrule height 0.7ex
			depth\dimexpr0.4pt-0.7ex\hskip3pt\kern0pt
			} Trained model\\
			\hskip3pt\raisebox{0.25ex}{\scalebox{0.75}{\color{datacolor}\textbullet}}\hskip4pt Data
		};
\end{tikzpicture}}

\begin{document}


\title[Fast Regression of the Tritium Breeding Ratio in Fusion Reactors]{Fast Regression of the
Tritium Breeding Ratio in Fusion Reactors}

\author{P~Mánek$^{1,2}$, G~Van Goffrier$^1$, V~Gopakumar$^3$, N~Nikolaou$^1$, J~Shimwell$^3$ and I~Waldmann$^1$}

\address{$^1$ Department of Physics and Astronomy, University College London, Gower Street, London WC1E~6BT, UK}
\address{$^2$ Institute of Experimental and Applied Physics, Czech Technical University, Husova 240/5, Prague 110~00, Czech Republic}
\address{$^3$ UK Atomic Energy Authority, Culham Science Centre, OX14~3DB Abingdon, UK}

\eads{\mailto{petr.manek.19@ucl.ac.uk}, \mailto{graham.vangoffrier.19@ucl.ac.uk}}

\begin{abstract}
	The tritium breeding ratio (TBR) is an essential quantity for the design of
	modern and next-generation D-T fueled nuclear fusion reactors. Representing the
	ratio between tritium fuel generated in breeding blankets and fuel consumed
	during reactor runtime, the TBR depends on reactor geometry and material
	properties in a complex manner. In this work, we explored the
	training of surrogate models to produce a cheap but high-quality approximation
	for a Monte Carlo TBR model in use at the UK Atomic Energy Authority. We
	investigated possibilities for dimensional reduction of its feature space, reviewed
	9~families of surrogate models for potential
	applicability, and performed hyperparameter optimisation. Here we present the
	performance and scaling properties of these
	models, the fastest of which, an artificial neural network,
	demonstrated~$R^2=\num{0.985}$ and a mean
	prediction time of~$\SI{0.898}{\micro\second}$, representing a relative speedup of $8\cdot 10^6$
	with respect to the expensive MC model. We further present a novel adaptive
	sampling algorithm, Quality-Adaptive Surrogate Sampling, capable
	of interfacing with any of the individually studied surrogates. Our preliminary
	testing on a toy TBR theory has demonstrated the efficacy of this algorithm for
	accelerating the surrogate modelling process.
\end{abstract}

\vspace{2ex}\\\submitto{Machine Learning: Science and Technology}
\maketitle
\ioptwocol


\section{Introduction}
\label{sec:introduction}
Surrogate models were developed to resolve computational limitations in the analysis of massive datasets by replacing a resource-expensive procedure with a much cheaper approximation
\cite{Sondergaard2003}. They are especially useful in applications where
numerous evaluations of an expensive procedure are required over the same or
similar domains, e.g.~in the parameter optimisation of a theoretical model. The
term ``metamodel'' proves especially meaningful in this case, when the surrogate
model approximates a computational process which is itself a model for a
(perhaps unknown) physical process~\cite{Myers2002}. There exists a spectrum
between ``physical'' surrogates which are constructed with some contextual
knowledge in hand, and ``empirical'' surrogates which are derived purely from
samples of the underlying expensive model.

In this work, we develop a family of empirical surrogate models for the tritium breeding
ratio (TBR) in an inertial confinement fusion (ICF) reactor. The expensive model
that our surrogate model approximates is a Monte Carlo (MC) neutronics
simulation, Paramak~\cite{paramak}, which returns a prediction of the TBR for a given
configuration of a spherical ICF reactor. Although more expensive 3D parametric models exist, we chose the Paramak simulation for its preferable speed in dataset generation in order to most fully demonstrate our methods. We quantify the success of several of our best-performing surrogate models by studying their accuracy and prediction time. We further propose an adaptive sampling algorithm (QASS) suitable for reducing the quantity of expensive samples needed to train our surrogate models.

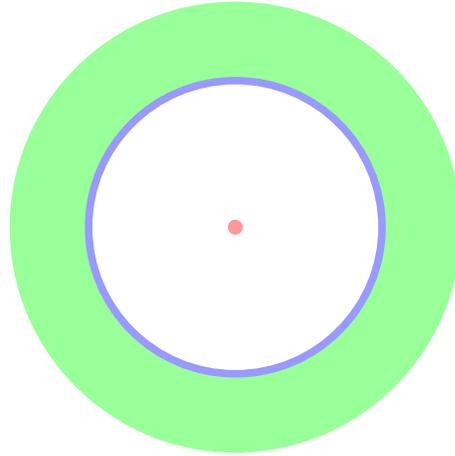
\begin{figure}[!ht]
  \centering
    \begin{tikzpicture}
    \fill[green!40!white]  (0,0) circle (3cm);
    \fill[blue!40!white]  (0,0) circle (2cm);
    \fill[white!40!white]  (0,0) circle (1.9cm);
    \fill[red!40!white]  (0,0) circle (0.1cm);
    \end{tikzpicture}

    \caption{Diagram of the simple sphere geometry (not to scale) where the
	blanket is \fcolorbox{white}{green}{\rule{0pt}{6pt}\rule{6pt}{0pt}}, the
first wall is \fcolorbox{white}{blue}{\rule{0pt}{6pt}\rule{6pt}{0pt}} and the
neutron point source is \fcolorbox{white}{red}{\rule{0pt}{6pt}\rule{6pt}{0pt}}.
Blanket and first wall thickness, as well as their material and structural
properties, are adjustable parameters of the simulation that are later optimized
(see~\Tref{tbl:params} for complete parameter listing).}
    \label{fig:model_diagram}
\end{figure}

Paramak facilitates simulation via an OpenMC neutronics workflow that is enclosed in
a portable Docker container, which conveniently exposes an HTTP API using the
Python~3 \texttt{flask} package. Within this setup, we employ a Muir energy distribution \cite{openmcmuir}\footnotemark[1]
around~\SI{14.06}{\mega\electronvolt} to approximate a Deuterium-Tritium (D-T)
plasma neutron source. As illustrated in~\Fref{fig:model_diagram}, the simulated reactor
geometry was made adjustable in order to study its influence on the TBR. Nuclear data for simulation were extracted from the following sources, in order of preference: FENDL 3.1d \cite{fendl31d}; JEFF 3.3 \cite{jeff33}; and ENDF/B-VII.1 \cite{2011ii}.
To maintain a model-agnostic approach, variance reduction (VR) techniques
were not used to accelerate the MC neutronics simulation~\cite{Kleijnen2013}.
It should be noted that depending on application, VR may constitute a viable
alternative to the presented work.

For the remainder of~\Sref{sec:introduction}, we will define the TBR and further motivate our
research. In~\Sref{sec:methodology} we will present our methodologies for the comparison
testing of a wide variety of surrogate modelling techniques, as well as defining an add-on adaptive sampling procedure QASS. After delivering the
results of these approaches in~\Sref{sec:results}, we will give our final conclusions and
recommendations in~\Sref{sec:conclusion}.

\footnotetext[1]{A bug in the Muir distribution involving erronous normal
sampling was recently uncovered (see
\url{https://github.com/openmc-dev/openmc/pull/1670}), but is disregarded in the present methods-focused work.}

\subsection{Problem Description}
\label{sec:problemdescription}

Nuclear fusion technology relies on the production and containment of an
extremely hot and dense plasma containing enriched Hydrogen isotopes. The current frontier generation of fusion reactors, such as the Joint European Torus (JET) and the
under-construction ITER, make
use of both tritium and deuterium fuel. While at least one deuterium atom occurs for every \num{5000} molecules of naturally-sourced water, and may be easily distilled, tritium is extremely rare in nature. Tritium may be produced indirectly through irradiation of heavy water
(\DDO) during nuclear fission, but only at very low rates which could
never sustain industrial-scale fusion power.

Modern D-T reactors rely on tritium breeding blankets, specialized
layers of material which partially line the reactor and produce tritium upon
neutron bombardment, e.g.~by 
\begin{eqnarray}
	\isotope[1][0]{n} + \hspace{3pt} \isotope[6][3]{Li} 
	&\longrightarrow \hspace{3pt} 
	\isotope[3][1]{T} + \hspace{3pt}\isotope[4][2]{He} \\
	\isotope[1][0]{n} + \hspace{3pt} \isotope[7][3]{Li} 
	&\longrightarrow \hspace{3pt} 
	\isotope[3][1]{T} + \hspace{3pt} \isotope[4][2]{He} + \hspace{3pt} \isotope[1][0]{n}.
\end{eqnarray}%
The TBR is defined as the ratio of tritium produced per source neutron, whose description in Paramak is facilitated by two classes of parameters
(exhaustively listed in~\Tref{tbl:params}). While the geometry of a given
reactor is described by continuous parameters, material selections are specified
by discrete categorical parameters. For all parameters, we have attempted to cover the full theoretical range of values even where those values are practically infeasible with current technology (e.g. packing fractions close to 1). Simulating broadly around typical values of parameters also improves the accuracy of the model nearer to typical values, and further aids in demonstrating the robustness of constructed models.

In our work, we set out to produce a fast TBR surrogate model, which takes the same input parameters as the MC model used in Paramak and approximates its output with the greatest achievable regression performance, while also minimising the required quantity of expensive-model samples needed for training. This represents a significant step forwards in computational fusion-reactor design, as any speed-up achieved in TBR evaluation directly informs a speed-up in numerical optimization of reactor parameters, although such optimization is beyond the scope of the present work.

\begin{table}[t]
	\setlength\tabcolsep{2pt}
	\renewcommand{\arraystretch}{0.95}
	\caption{\label{tbl:params}Input parameters supplied to Paramak and surrogates in alphabetical order. Groups of fractions marked\textsuperscript{\textdagger
		\textdaggerdbl} are independently required to sum to 1.}
	\begin{indented}
	\item[]
		\begin{tabular}{l|ll}
		\toprule
		{} & Parameter name (abbreviation) & Domain\\
		\midrule
		\parbox[t]{2mm}{\hspace{-2pt}\multirow{12}{*}{\rotatebox[origin=c]{90}{Blanket}}}
		   & Breeder fraction\textsuperscript{\textdagger} & $[0,1]$\\
		   & Breeder \isotope[6]{Li} enrichment fraction & $[0,1]$\\
		   & Breeder material (BBM) & $\{\text{Li}_2\text{TiO}_3, \text{Li}_4\text{SiO}_4\}$\\
		   & Breeder packing fraction & $[0,1]$\\
		   & Coolant fraction\textsuperscript{\textdagger} & $[0,1]$\\
		   & Coolant material (BCM) & $\{\text{D}_2\text{O}, \text{H}_2\text{O}, \text{He}\}$\\
		   & Multiplier fraction\textsuperscript{\textdagger} & $[0,1]$\\
		   & Multiplier material (BMM) & $\{\text{Be}, \text{Be}_{12}\text{Ti}\}$\\
		   & Multiplier packing fraction & $[0,1]$\\
		   & Structural fraction\textsuperscript{\textdagger} (BSM) & $[0,1]$\\
		   & Structural material & $\{\text{SiC}, \text{eurofer}\}$\\
		   & Thickness & $[0,500]\text{ cm}$\\
		\midrule
		\parbox[t]{2mm}{\hspace{-2pt}\multirow{6}{*}{\rotatebox[origin=c]{90}{First wall}}}
		   & Armour fraction\textsuperscript{\textdaggerdbl} & $[0,1]$\\
		   & Coolant fraction\textsuperscript{\textdaggerdbl} & $[0,1]$\\
		   & Coolant material (FCM) & $\{\text{D}_2\text{O}, \text{H}_2\text{O}, \text{He}\}$\\
		   & Structural fraction\textsuperscript{\textdaggerdbl} & $[0,1]$\\
		   & Structural material (FSM) & $\{\text{SiC}, \text{eurofer}\}$\\
		   & Thickness & $[0,20]\text{ cm}$\\
		\bottomrule
		\end{tabular}
	\end{indented}
\end{table}

\section{Methodology}
\label{sec:methodology}
Labeling the expensive Paramak model $f(x)$, a surrogate model is a function
$\hat{f}(x)$ such that $f(x)$ and $\hat{f}(x)$ minimize a selected dissimilarity
metric. In order to be considered \textit{viable}, $\hat{f}(x)$ is required to
achieve an expected evaluation time lower than that of~$f(x)$. In this work, we
consider two methods of producing viable surrogates: (1)~a conventional decoupled
approach, which evaluates $f(x)$ on a set of uniformly-random samples and
trains surrogates in a supervised scheme, and (2)~an adaptive approach, which attempts to
compensate for localized regression performance insufficiencies by interleaving
multiple epochs of sampling and training. Several high-accuracy and deployment-ready surrogate models are developed using the decoupled approach, and their performance characterized numerically, while the adaptive approach is studied exclusively as a proof-of-concept.

\begin{table}[t]
	\setlength\tabcolsep{1pt}
	\renewcommand{\arraystretch}{0.95}
	\caption{\label{tbl:surrogates}Considered surrogate model families, their
		selected abbreviations and implementations. $\mathcal{H}$~denotes the
		set of hyperparameters, family-dependent priors that control the
		learning process, and are tuned separately. Families with fewer
		hyperparameter represents a smaller surogate domain to explore.}
	\begin{indented}
	\item[]
		\begin{tabular}{lllr}
		\toprule
		Surrogate family & Abbr. & Impl. & $|\mathcal{H}|$ \\
		\midrule
		Support vector machines~\cite{fan2008liblinear}	& SVM & SciKit~\cite{scikit-learn} & 3 \\
		Gradient boosted trees~\cite{friedman2001greedy,friedman1999stochastic,hastie2009elements}	& GBT & SciKit & 11 \\
		Extremely randomized trees~\cite{geurts2006extremely}	& ERT & SciKit & 7 \\
		AdaBoosted decision trees$^\text{a}$~\cite{drucker1997improving}	& ABT & SciKit & 3 \\
		Gaussian process regression~\cite{williams2006gaussian}	& GPR & SciKit & 2 \\
		$k$ nearest neighbours	& KNN & SciKit & 3 \\
		Artificial neural networks	& ANN & Keras~\cite{chollet2015keras} & 2 \\
		Inverse distance weighing~\cite{shepard1968two} & IDW & SMT~\cite{SMT2019} & 1 \\
		Radial basis functions & RBF & SMT & 3 \\
		\bottomrule
		\end{tabular}\\%
		{\footnotesize $^\text{a}$Note that ABTs can be viewed as a subclass of GBTs.}
	\end{indented}
\end{table}

\begin{table*}[t]
	\renewcommand{\arraystretch}{0.95}
	\caption{\label{tbl:metrics}Metrics recorded in experiments. In
	formulations, we work with a training set of size $N_0$ and a test set of
size $N$, values $y^{(i)}=f(x^{(i)})$ and $\hat{y}^{(i)}=\hat{f}(x^{(i)})$
denote images of the $i$th testing sample in Paramak and the surrogate
respectively. The mean $\overline{y}=N^{-1}\sum_{i=1}^N y^{(i)}$ and $P$ is the
number of input features.}
	\begin{indented}
	\item[]
		\begin{tabularx}{\textwidth}{Xrl}
		\toprule
		Regression performance metrics& Notation	& Mathematical formulation\\
		\midrule
		Mean absolute error	& MAE & $N^{-1}\sum_{i=1}^N |y^{(i)}-\hat{y}^{(i)}|$ \\
		Standard deviation of error & $S$	& $\text{StdDev}_{i=1}^N\left\{ |y^{(i)} -
		\hat{y}^{(i)}| \right\} $ \\
			Coefficient of determination & $R^2$	& $1-\sum_{i=1}^N
			\left(y^{(i)}-\hat{y}^{(i)} \right)^2\left[\sum_{i=1}^N \left(
			y^{(i)}-\overline{y} \right)^2\right]^{-1} $ \\
			Adjusted $R^2$ & $R^2_\text{adj.}$	& $1-(1-R^2)(N-1)(N-P-1)^{-1}$ \\
		\midrule
		Computational complexity metrics	& {}	& {} \\
		\midrule
		Mean training time & $\overline{t}_{\text{trn.}}$	& $(\text{wall training time of
		$\hat{f}(x)$})N_0^{-1}$  \\
			Mean prediction time & $\overline{t}_{\text{pred.}}$	& $(\text{wall prediction time of
			$\hat{f}(x)$})N^{-1}$ \\
				Relative speedup & $\omega$	& $(\text{wall evaluation$^\text{b}$ time of $f(x)$})
				(N\overline{t}_{\text{pred.}})^{-1}$ \\
		\bottomrule
		\end{tabularx}\\%
		{\footnotesize $^\text{b}$This corresponds to evaluation of Paramak
		 on all points of the test set. In surrogates, the equivalent
		time period is referred to as the ``prediction time.''}
	\end{indented}
\end{table*}

We selected several state-of-the-art regression algorithms to perform
surrogate training on sampled point sets. Listed in~\Tref{tbl:surrogates}, these
implementations define nine surrogate families which are detailed in~\Sref{sec:results}.
We note that each presented algorithm defines hyperparameters that may influence its
performance. Their problem-specific optimal values are searched within the scope
of this work, in particular in Experiments~1 \&~2 that are outlined
in~\Sref{sec:experiment-methodology}.

To compare the quality of the produced surrogates, we define a variety of metrics listed
in~\Tref{tbl:metrics}. For regression performance analysis, we include a
selection of absolute metrics (MAE, $S$) to assess the models' approximation capability
and to set practical bounds on the expected uncertainty of their predictions. In addition, we also track
relative measures ($R^2$, $R^2_\text{adj.}$) that are better-suited for comparison between this work and others, as
they are invariant with respect to the selected domain and image space.
For analysis of computational complexity, surrogates are assessed in terms of wall
time (captured by the Python~3 \texttt{time} package). This is motivated by common practical use-cases of our work, where surrogate models are trained as replacements for
Paramak. All times reported (training, test, evaluation) are
normalized by the corresponding dataset size, i.e.~correspond to ``time to
process a single datapoint.''

Even though some surrogates support acceleration by means of parallelization, we
used non-parallelized implementations. The only exception to this is the ANN family,
which requires a considerable amount of processing power for training on
conventional CPU architectures. Lastly, to prevent undesirable bias by training
set selection, all reported metrics are obtained via five-fold cross-validation.
In this setting, a sample set is uniformly divided into five disjoint folds, each of which
is used as a test set for models trained on the remaining four. Having repeated the
same experiment for each division, the overall value of individual metrics is reported in terms of their mean and standard deviation over all folds.

\subsection{Decoupled Approach}\label{sec:experiment-methodology}

Experiments related to the decoupled approach are organized in four parts,
further described in this section. In summary, we aim to optimize the hyperparameters of
each surrogate family separately, and later compare the best results between
families.

The objective of Experiment~1 is to simplify the regression task for
surrogates prone to suboptimal performance in discrete spaces.
To this end, training points are filtered to a single selected discrete feature
assignment, and surrogates are trained only on the remaining continuous features.
This is repeated several times to explore variances in behavior,
particularly in four distinct assignments that are obtained by setting blanket and
first wall coolant materials to one
of:~$\{\text{H\textsubscript{2}O},\isotope{He}\}$.
Experiment~2 conventionally measures surrogate performance on the full feature
space without any parameter restrictions. In both experiments, hyperparameter tuning is
facilitated by Bayesian optimisation~\cite{movckus1975bayesian}, where we select the
hyperparameter configuration that produces the model that maximizes $R^2$. The
process is terminated after 1000~iterations or two days, whichever condition is satisfied first.
The results of Experiments~1 \&~2 are depicted
in Figures~\ref{fig:exp1-time-vs-reg} \&~\ref{fig:exp2-time-vs-reg}
respectively, and described in~\Sref{sec:res-exp12}.

In Experiment~3, the twenty best-performing hyperparameter configurations
for each model family are used to train surrogates on sets of various sizes to
investigate their scaling properties. In particular, we track the metrics
from~\Tref{tbl:metrics} as functions of training set size (1, 2, 5, 10, 12, 15 and 20 thousands
of samples) individually for each family. This allows their comparison based on
observed trends, and estimation of optimal training set sizes.
The results of this experiment are shown in~\Fref{fig:scaling} and discussed
in~\Sref{sec:res-exp3}.

Finally, Experiment~4 aims
to produce surrogates suitable for practical use by retraining selected
well-scaling instances on large training sets.
The results of this process are displayed in~\Fref{fig:reg-performance} and
in~\Tref{tbl:exp4-detailed-results}, and summarized in~\Sref{sec:res-exp4}.

\subsection{Adaptive Approach}\label{sec:adaptive}

Adaptive sampling techniques appear frequently in the literature and have been
specialized for surrogate modelling, where precision is implicitly limited by the quantity of training samples which are available from the expensive model. Garud's~\cite{Garud2016} ``Smart Sampling Algorithm'' achieved notable success by incorporating surrogate quality and
crowding distance scoring to identify optimal new samples, but was only tested
on a single-parameter domain. We theorized that a nondeterministic sample
generation approach, built around Markov Chain Monte Carlo methods (MCMC), would
fare better for high-dimensional models by more thoroughly exploring all local
optima in the feature space. MCMC produces each sample point according to a jump step drawn from a shared proposal
distribution. These sample points will converge to a desired posterior
distribution, so long as the acceptance probability meets certain statistical criteria (see~\cite{Zhou2018} for a review).

\begin{figure}
	\centering
	\hspace*{-5pt}\includegraphics[width=1.28\linewidth]{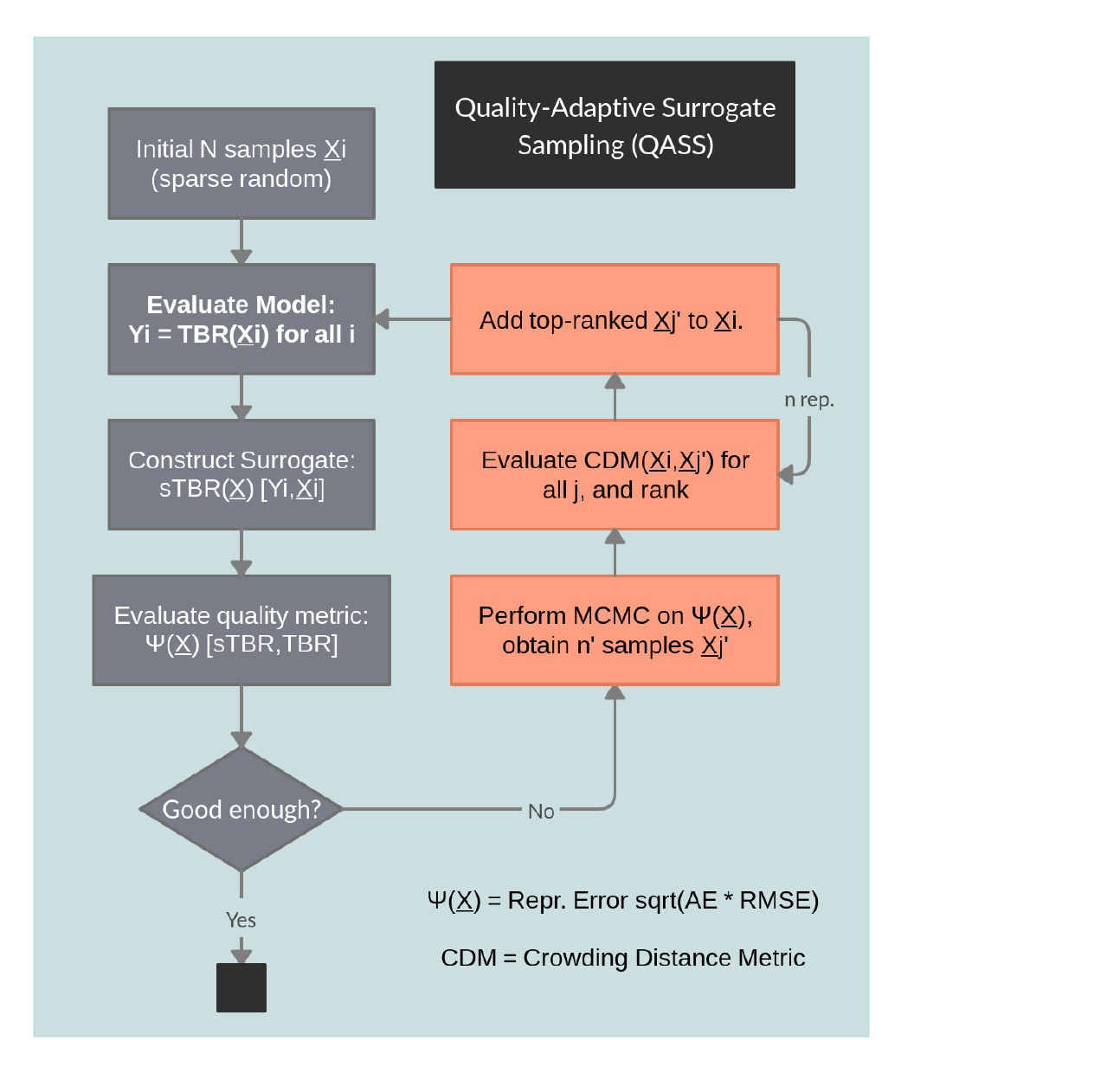}
	\caption{\label{fig:qassplan}Schematic of QASS algorithm}
\end{figure}

Many researchers have embedded surrogate methods into MCMC strategies for
parameter optimisation~\cite{Zhang2020,Gong2017}, in particular the ASMO-PODE
algorithm~\cite{Ginting2011} which makes use of MCMC-based adaptive sampling. Our approach draws inspiration from ASMO-PODE, but instead uses MCMC to generate samples
which increase surrogate precision throughout the entire parameter space.

We designed the Quality-Adaptive Surrogate Sampling algorithm (QASS, \Fref{fig:qassplan}) to iteratively increment the training/test set with sample
points which maximize surrogate error and minimize a crowding distance metric
(CDM)~\cite{Solonen2012} in feature space. Error maximization is desirable for these sample points because it identifies regions of parameter space where the surrogate most needs to be improved. On each iteration following an initial training of the surrogate on $N$ uniformly random samples, the surrogate was trained and absolute error calculated. MCMC was then performed to sample the error function generated by performing nearest-neighbor interpolation on these test error points. The resultant samples were culled by $50\%$ according to the CDM, and then the $n$ highest-error candidates were selected for reintegration with the training/test set, beginning another training epoch. Validation was also performed during each iteration on independent, uniformly-random sample sets.

\section{Results}
\label{sec:results}
\subsection{Decoupled Approach}
\label{sec:modelres}

\subsubsection{Hyperparameter Tuning}
\label{sec:res-exp12}

\begin{figure}
	\centering
	\hspace*{-0.05\columnwidth}
	\includegraphics[trim=10pt 50pt 10pt 10pt,clip,width=1.1\linewidth]{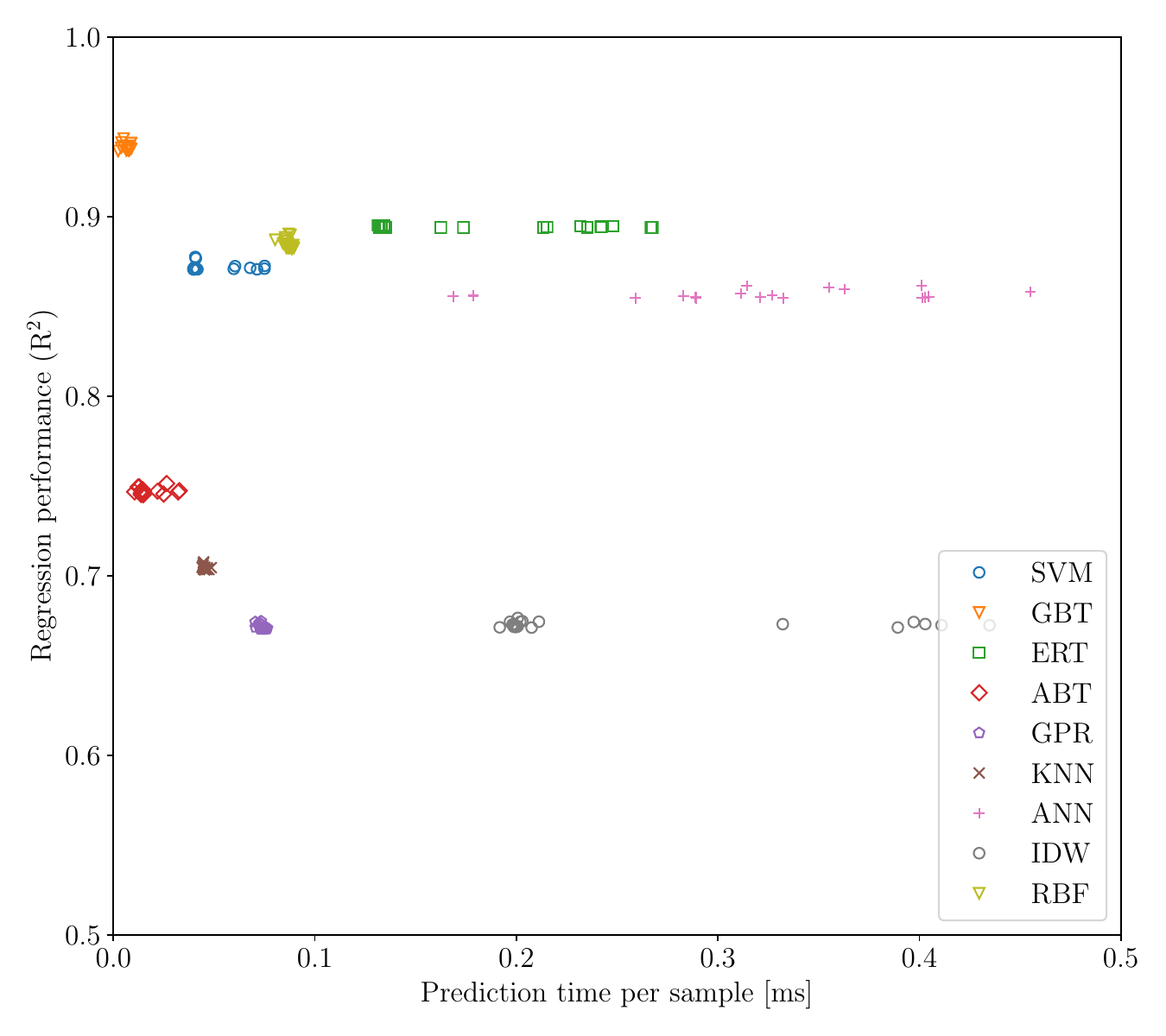}

	\vfill

	\hspace*{-0.05\columnwidth}
	\includegraphics[trim=10pt 50pt 10pt 10pt,clip,width=1.1\linewidth]{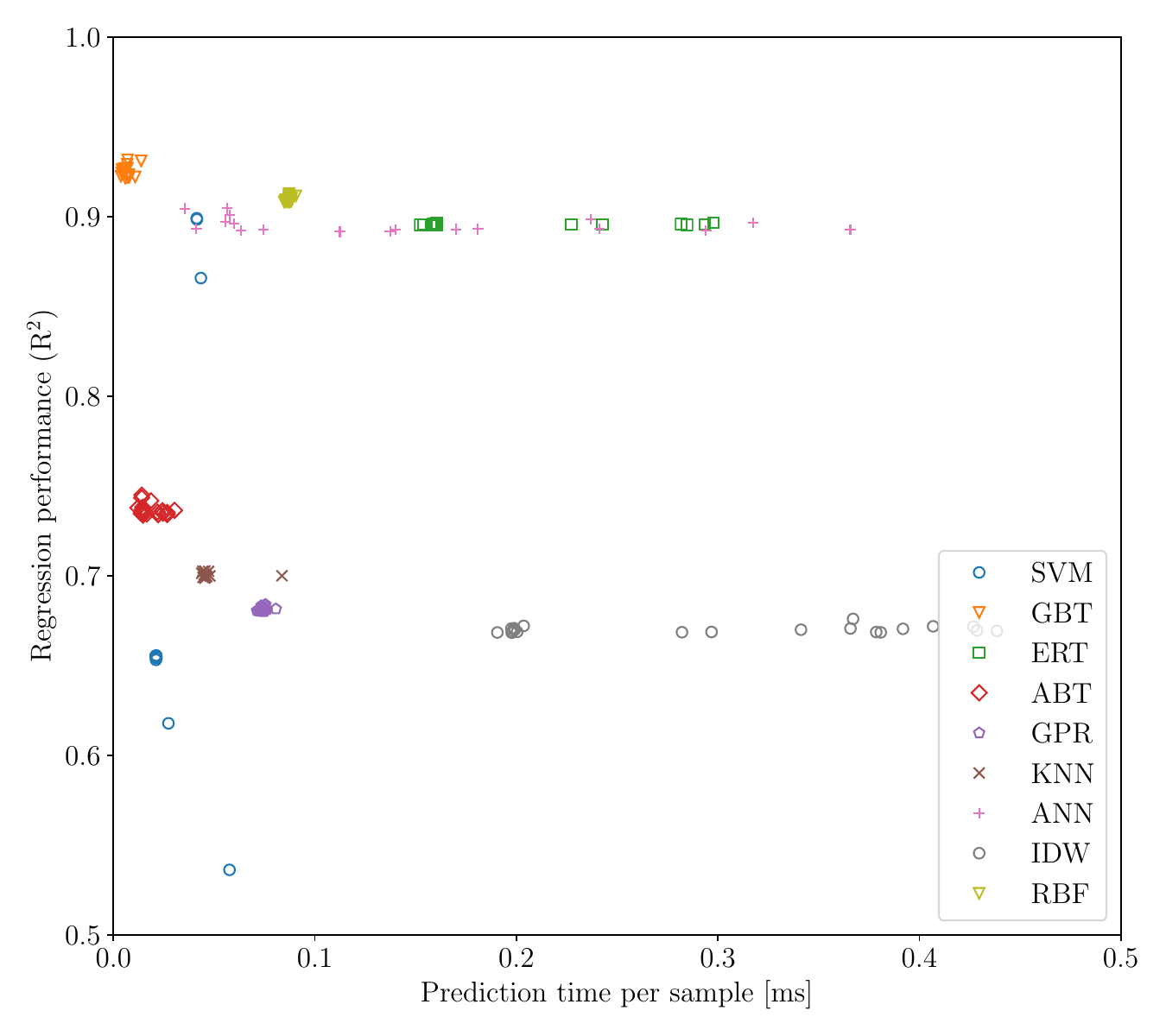}

	\vfill

	\hspace*{-0.05\columnwidth}
	\includegraphics[trim=10pt 15pt 10pt 10pt,clip,width=1.1\linewidth]{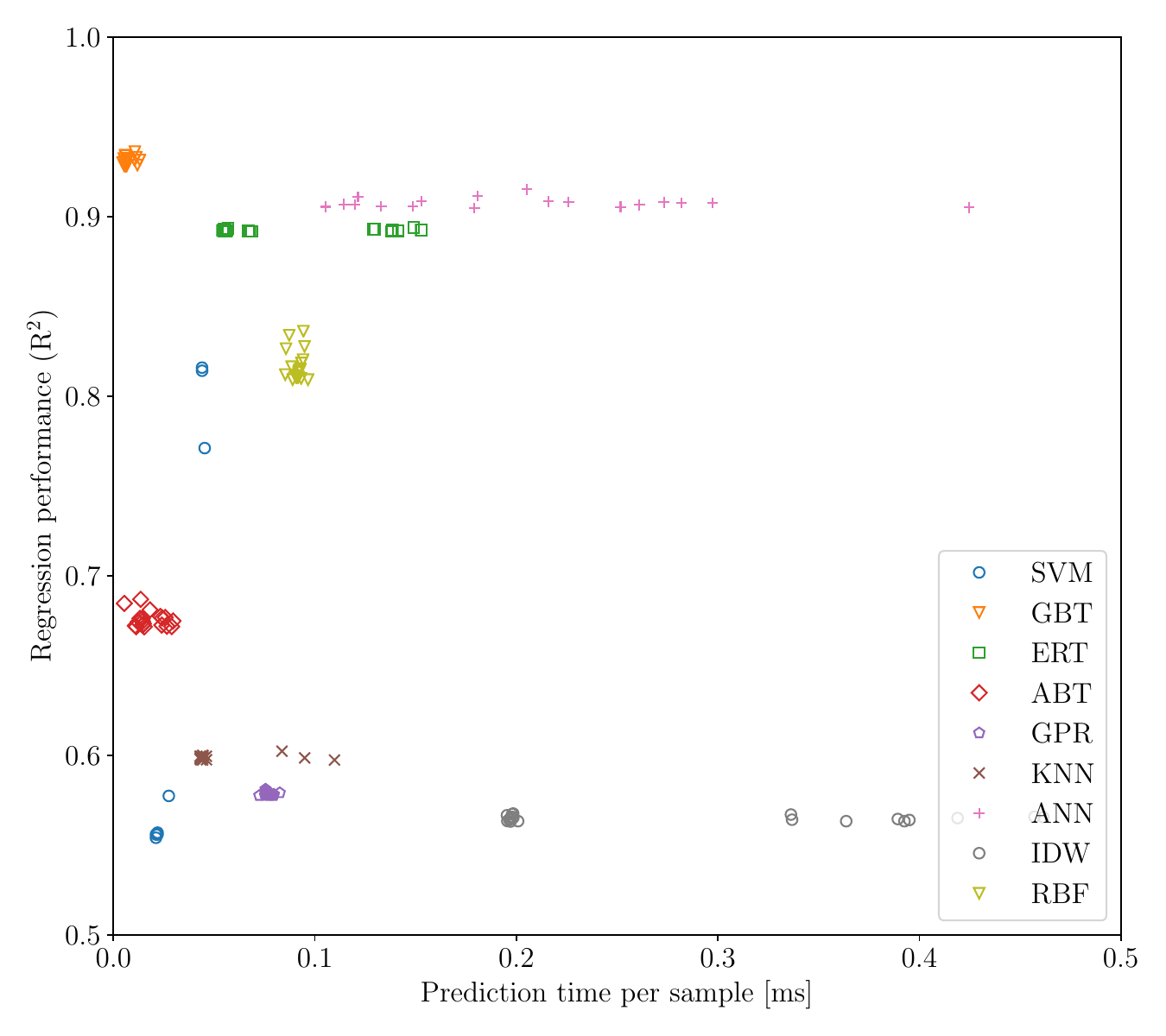}

	\caption{\label{fig:exp1-time-vs-reg}Experiment~1 results. 20~best surrogates per each considered family, plotted in
		terms of~$\overline{t}_{\text{pred.}}$ and~$R^2$ with 3~selected slices out
	of~4 (defined in~\Tref{tbl:slices}).}
\end{figure}

The results displayed in~\Fref{fig:exp1-time-vs-reg} indicate that in the first,
simplified case GBTs clearly appear to be the most accurate as well as the
fastest surrogate family in terms of mean prediction time. Following that, we
note that ERTs, SVMs and ANNs also achieved satisfactory results with respect to
both examined metrics. In addition, prediction times of GBTs and SVMs show
relatively lower variance than those of ERTs and ANNs. Even though the remainder
of tested surrogate families do not exhibit prohibitive complexity, their
regression performance fall below the average.

\begin{table}[h]
	\centering
	\setlength\tabcolsep{5pt}
	\renewcommand{\arraystretch}{0.95}
	{\footnotesize
		\begin{tabular}{llllll}
		\toprule
		BBM & BCM & BMM & BSM & FCM & FSM \\
		\midrule
		$\text{Li}_4\text{SiO}_4$ & $\text{H}_2\text{O}$ & $\text{Be}_{12}\text{Ti}$ & eurofer & $\text{H}_2\text{O}$ & eurofer\\
		$\text{Li}_4\text{SiO}_4$ & He & $\text{Be}_{12}\text{Ti}$ & eurofer & $\text{H}_2\text{O}$ & eurofer\\
		$\text{Li}_4\text{SiO}_4$ & $\text{H}_2\text{O}$ & $\text{Be}_{12}\text{Ti}$ & eurofer & He & eurofer\\
		$\text{Li}_4\text{SiO}_4$ & He & $\text{Be}_{12}\text{Ti}$ & eurofer & He & eurofer\\
		\bottomrule
		\end{tabular}
	}
	\caption{Slices 1-4 of the domain space (discrete parameter
	assignments) explored in Experiment~1. Columns correspond to abbreviated
	parameter names listed in~\Tref{tbl:params}.}
	\label{tbl:slices}
\end{table}

\begin{figure}
	\centering

	\hspace*{-0.1\columnwidth}
	\includegraphics[trim=10pt 15pt 10pt 10pt,clip,width=1.1\linewidth]{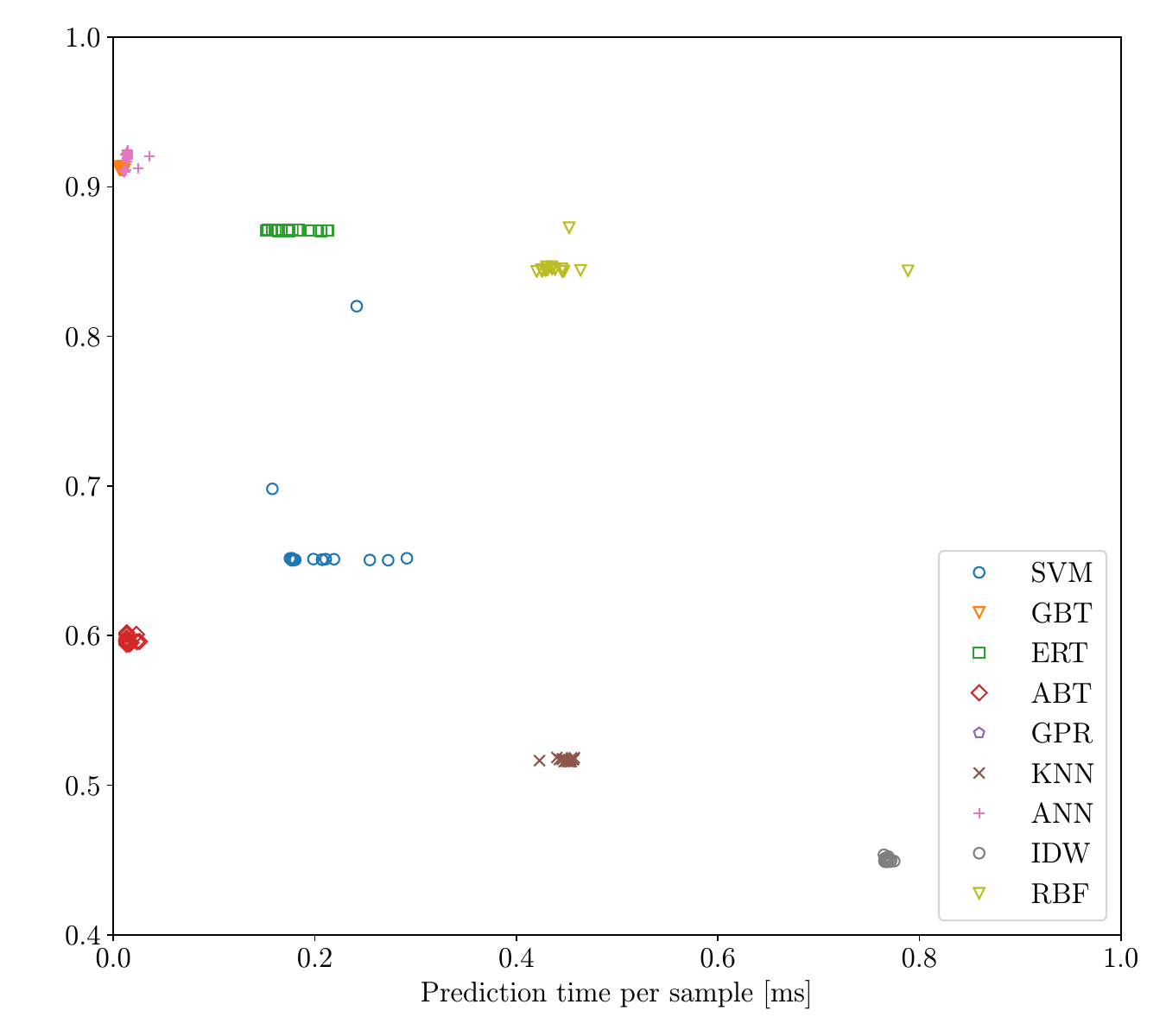}
	\caption{\label{fig:exp2-time-vs-reg}Experiment~2 results, plotted analogously
	to~\Fref{fig:exp1-time-vs-reg}.}
\end{figure}

Comparing these results with those of the second, unrestricted experiment (shown
in~\Fref{fig:exp2-time-vs-reg}), we observe that many surrogate families
consistently underperform. The least affected models appear to be GBTs, ANNs and
ERTs, which are known to be capable of capturing relationships involving mixed
feature types that were deliberately withheld in the first experiment. With only
negligible differences, the first two of these families appear to be tied for
the best performance as well as the shortest prediction time. We observe that
ERTs and RBFs also demonstrated satisfactory results, clearly outperforming the
remaining surrogates in terms of regression performance, and in some cases also prediction time.

Following both hyperparameter tuning experiments, we conclude that while domain
restrictions employed in the first case have proven effective in improving the
regression performance of some methods, their performance fluctuates considerably
depending on the selected slices. For instance, the variance in SVM performance
in slice 1 is much lower than in slices 2-3, and both KNNs and ABTs perform much
better in slices 1-2 than in slice 3. Furthermore, in all instances the best
results are achieved by families of surrogates that do not benefit from this
restriction (GBTs, ANNs, ERTs).

\subsubsection{Scaling Benchmark}
\label{sec:res-exp3}

\begin{figure}
	\centering
	\hspace*{-0.1\columnwidth}
	\includegraphics[trim=10pt 50pt 10pt 10pt,clip,width=1\columnwidth]{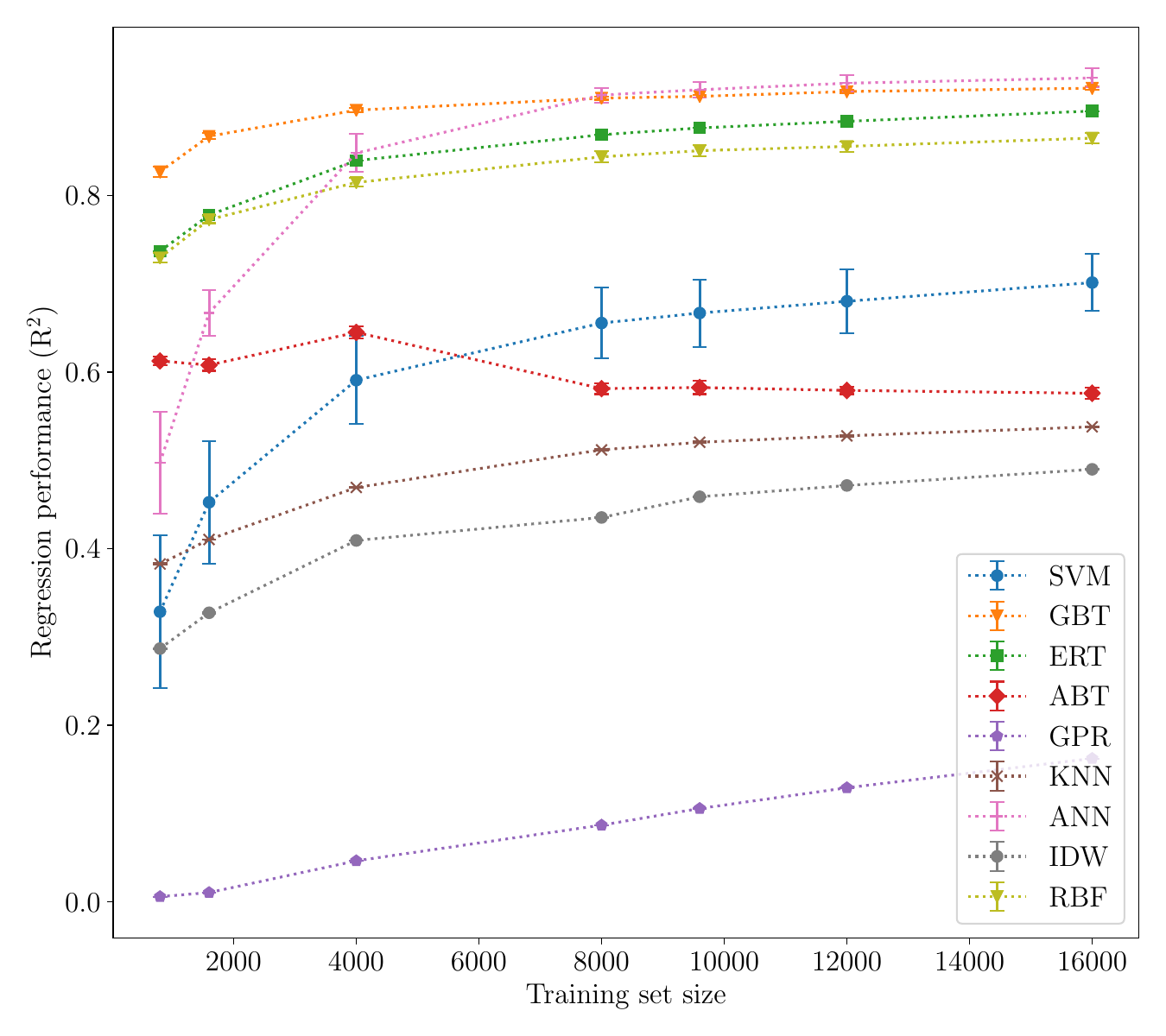}
	
	\vfill

	\hspace*{-0.1\columnwidth}
	\includegraphics[trim=5pt 50pt 10pt 10pt,clip,width=1\columnwidth]{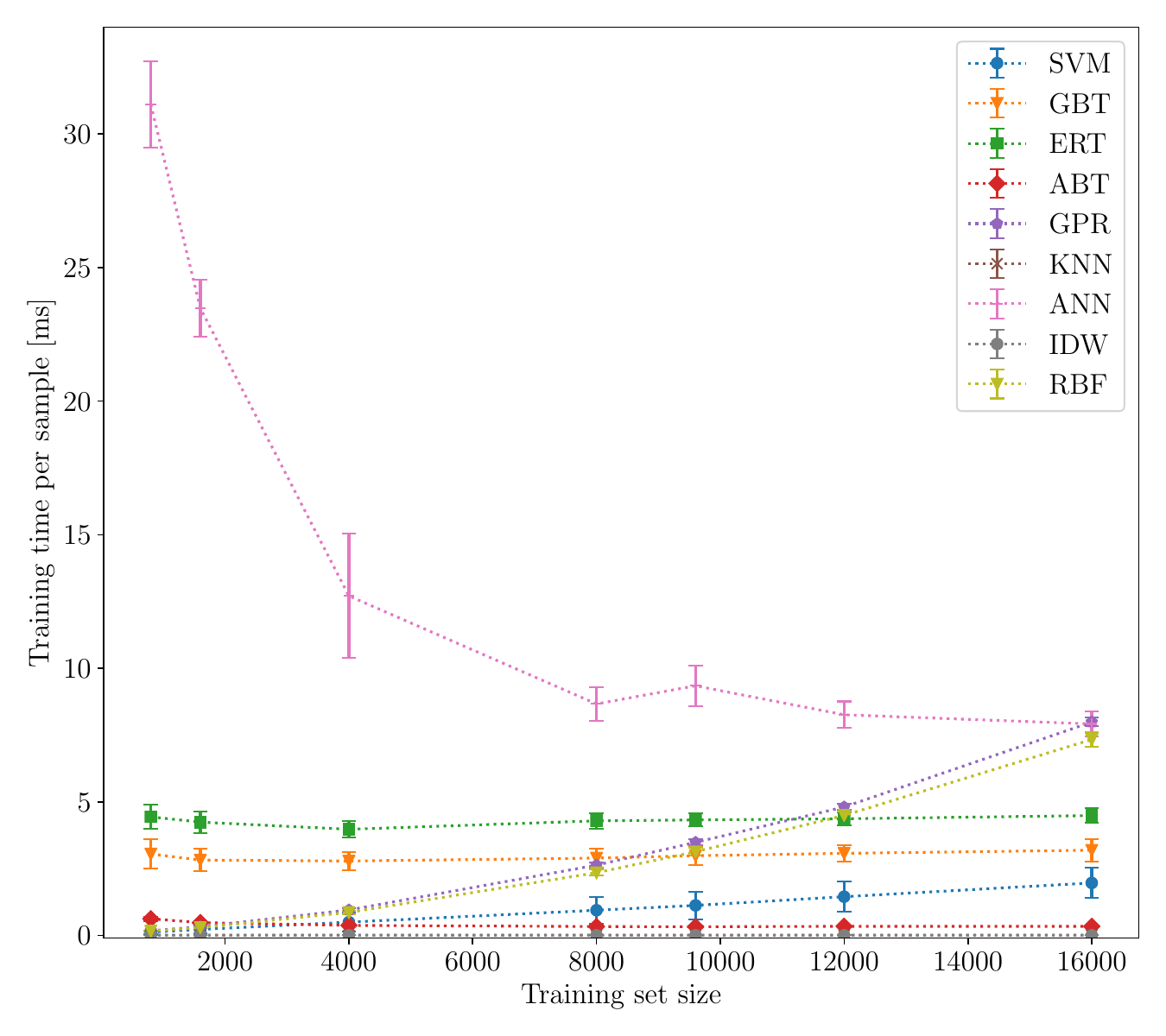}
	
	\vfill

	\hspace*{-0.1\columnwidth}
	\includegraphics[trim=0pt 15pt 10pt 15pt,clip,width=1\columnwidth]{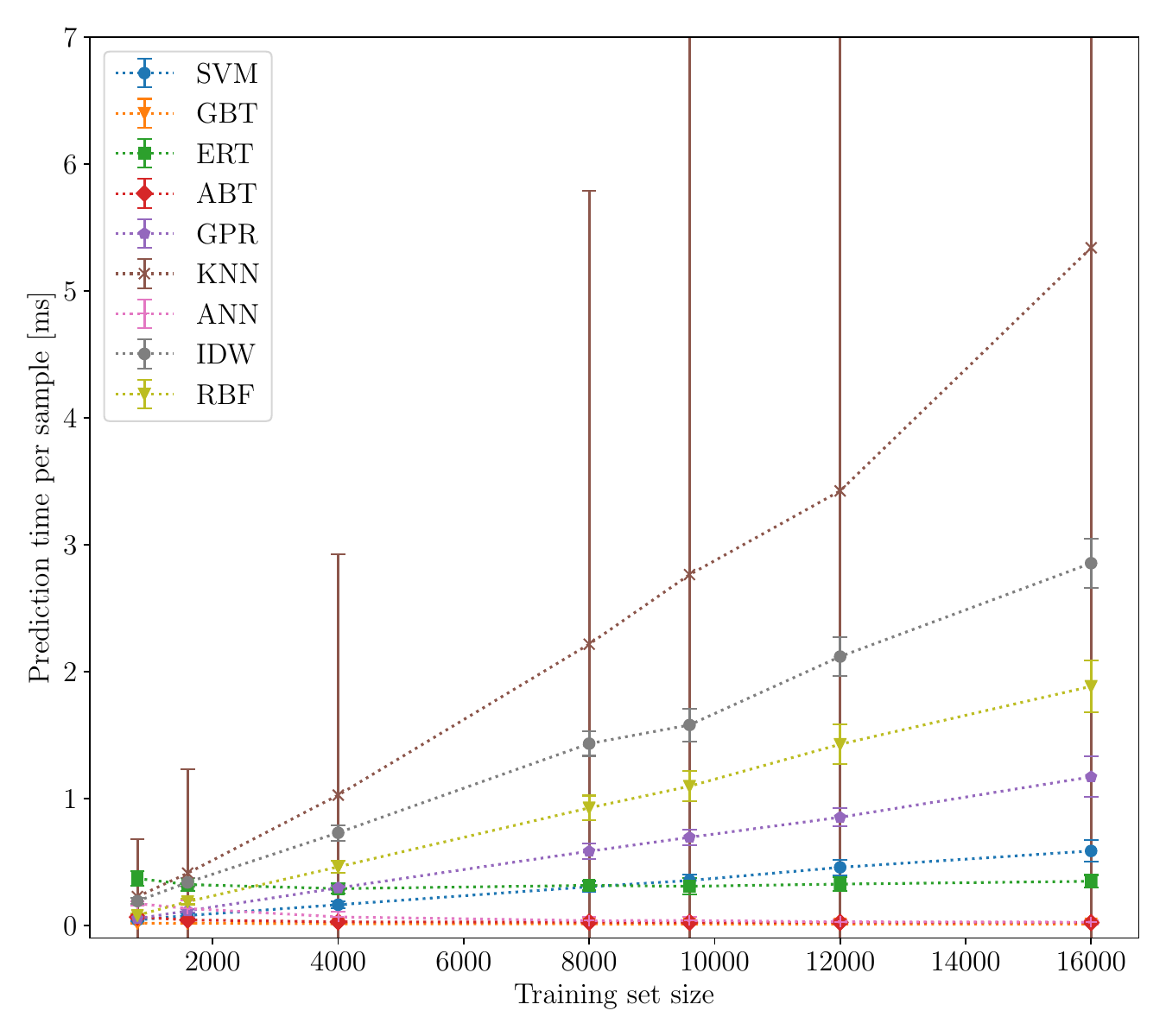}
	\caption{Experiment~3 results, displayed as a function of $N_0$. From top to bottom, $R^2$,
	$\overline{t}_{\text{trn.}}$, $\overline{t}_{\text{pred.}}$.}
	\label{fig:scaling}
\end{figure}

The results shown in~\Fref{fig:scaling} suggest that in terms of regression
performance the most accurate families from the previous experiments
consistently maintain their relative advantage over others, even as
more training points are introduced. While such families achieve nearly comparable
performance on the largest dataset, in the opposite case tree-based ensemble
approaches clearly outperform ANNs. For instance, GBTs achieve
$\text{MAE}=\num{0.107}$, nearly half of the $\text{MAE}=\num{0.186}$
achieved by ANNs, representing a clear benefit given vastly disparate training
and prediction times. This trend continues for set sizes up to~\num{6000}.

Consistent with our expectations, the shortest training times were achieved by
instance-based learning methods (KNN, IDW) that
are trained trivially at the expense of increased lookup complexity later during prediction.
Furthermore, we observe that the majority of tree-based ensemble algorithms also perform
and scale well, unlike RBFs and GPR which appear to behave superlinearly. We note that ANNs,
which are the only family to utilize parallelisation during training, show an
inverse scaling characteristic. We suspect that this effect may be caused
by a constant multi-threading overhead that dominates the training process
on relatively small sets.

Finally, all tested families with the exception of previously mentioned instance-based
models offered desirable prediction times. Analogous to previous experiments,
GBTs, ABTs and ANNs appeared to be tied, as they not only exhibited
comparable times but also similar scaling slopes. After those, we note a
clear hierarchy of ERTs, SVMs, GPR and RBFs, trailed by IDW and KNNs.

\subsubsection{Model Comparison}
\label{sec:res-exp4}
\begin{figure}
	\centering

	\hspace*{-0.06\linewidth}
	\begin{minipage}{0.53\linewidth}
		\centering
		\surrogateplot{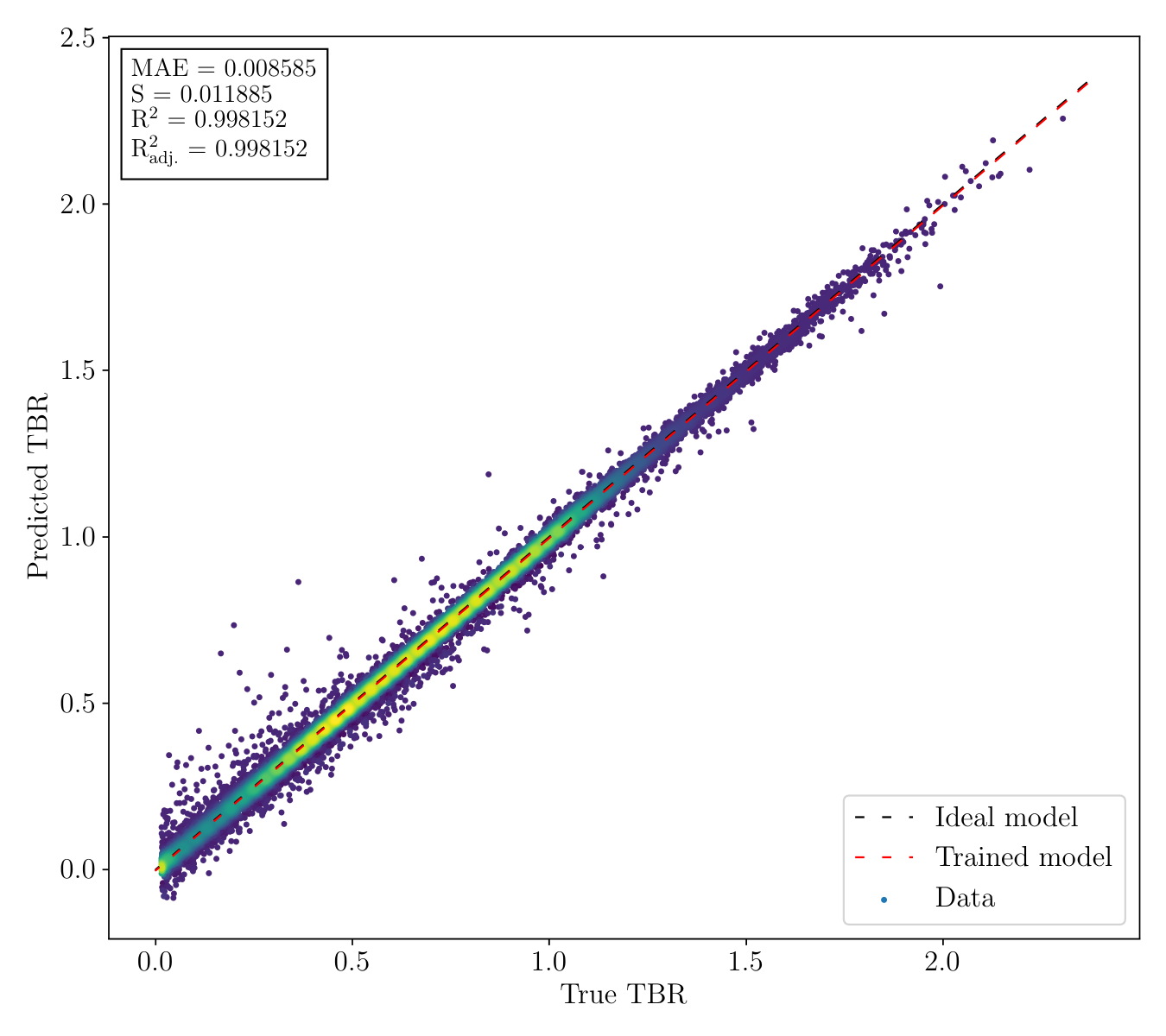}{0.008585}{0.011885}{0.998152}{0.998152}{3.79}
	\end{minipage}%
	\begin{minipage}{0.53\linewidth}
		\centering
		\surrogateplot{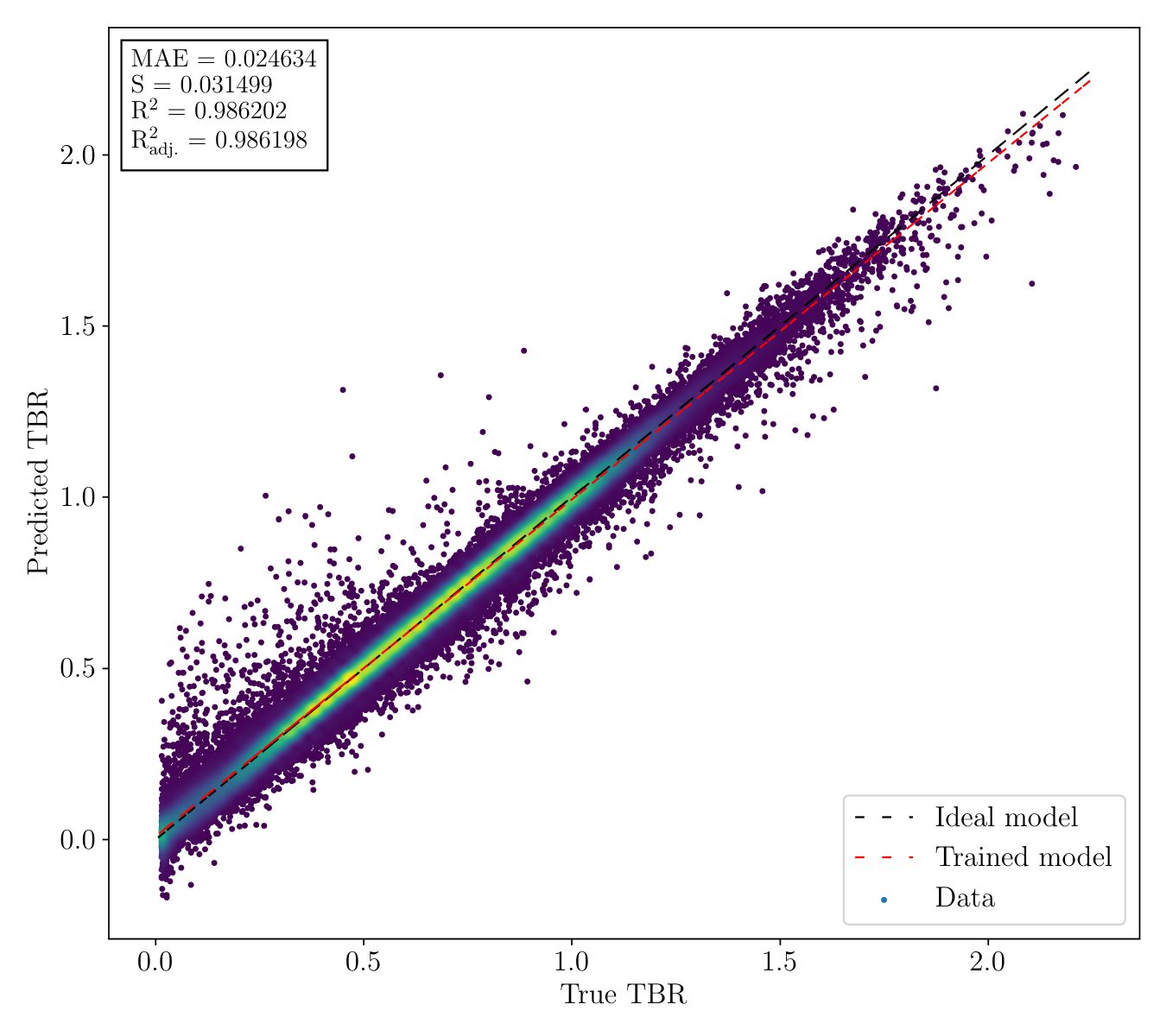}{0.024634}{0.031499}{0.986202}{0.986198}{3.82}
	\end{minipage}

	\hspace*{-0.06\linewidth}
	\begin{minipage}{0.53\linewidth}
		\centering
		\surrogateplot{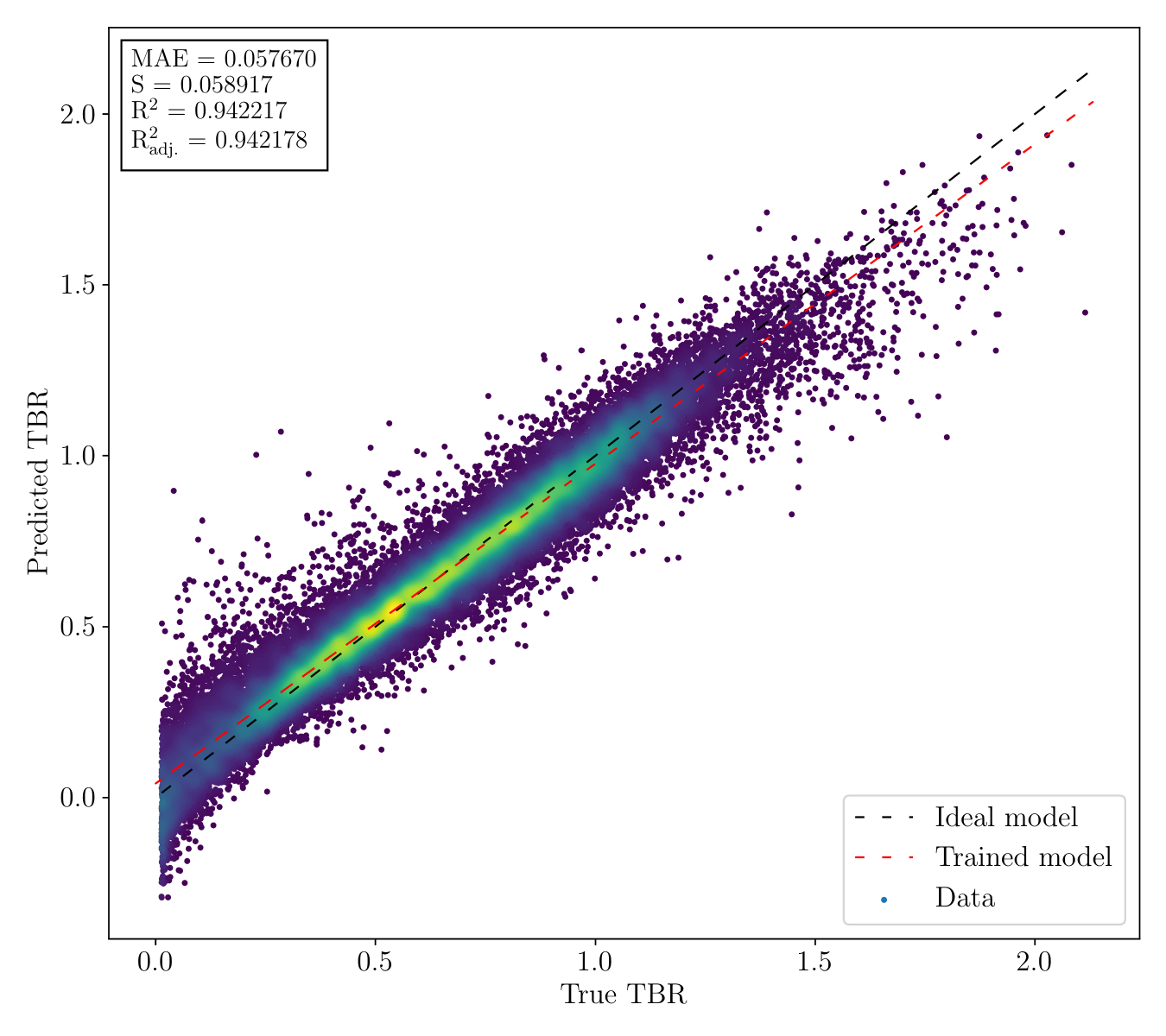}{0.057670}{0.058917}{0.942217}{0.942178}{3.82}
	\end{minipage}%
	\begin{minipage}{0.53\linewidth}
		\centering
		\surrogateplot{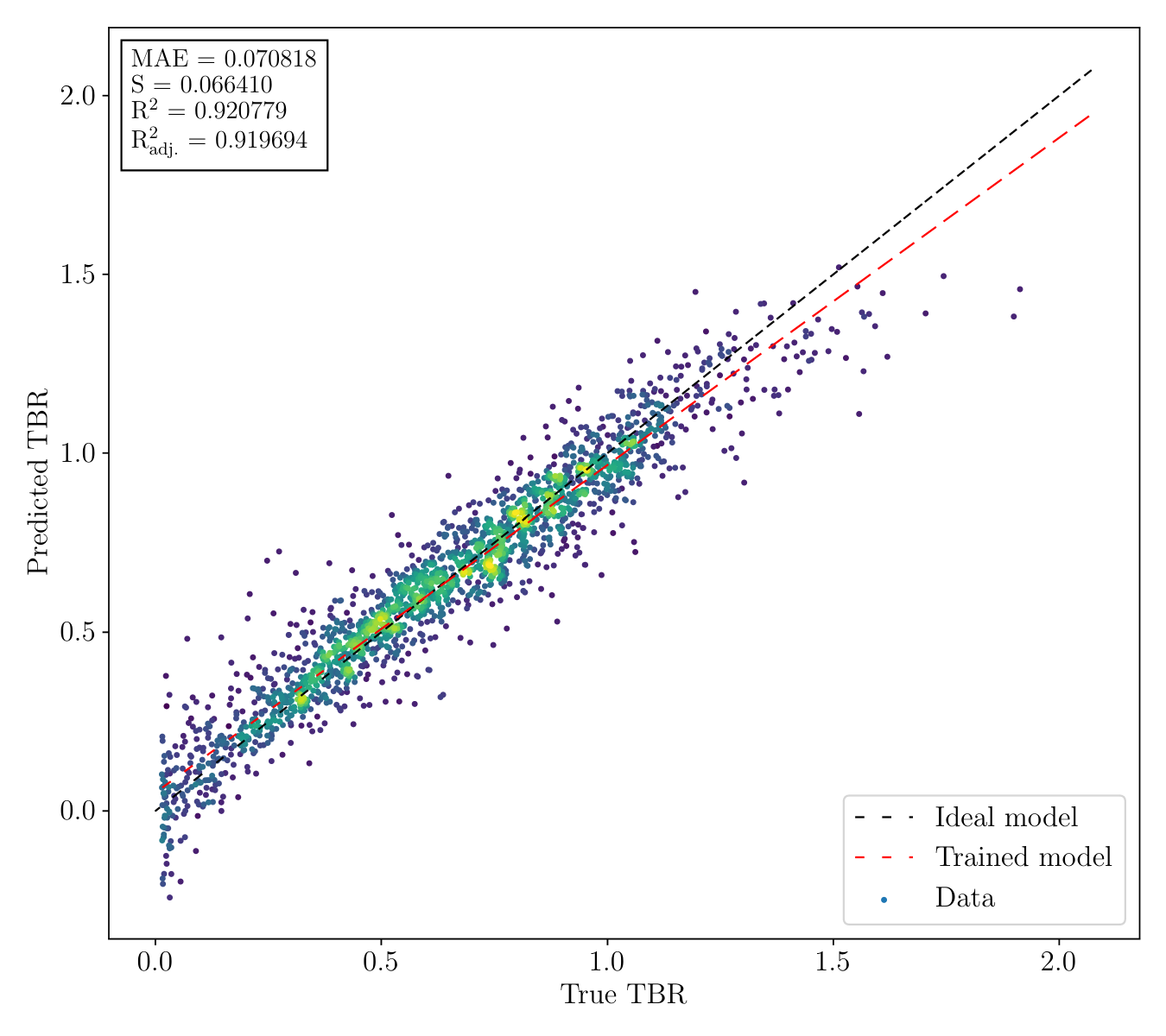}{0.070818}{0.066410}{0.920779}{0.919694}{3.82}
	\end{minipage}

	\hspace*{-0.06\linewidth}
	\begin{minipage}{0.53\linewidth}
		\centering
		\surrogateplot{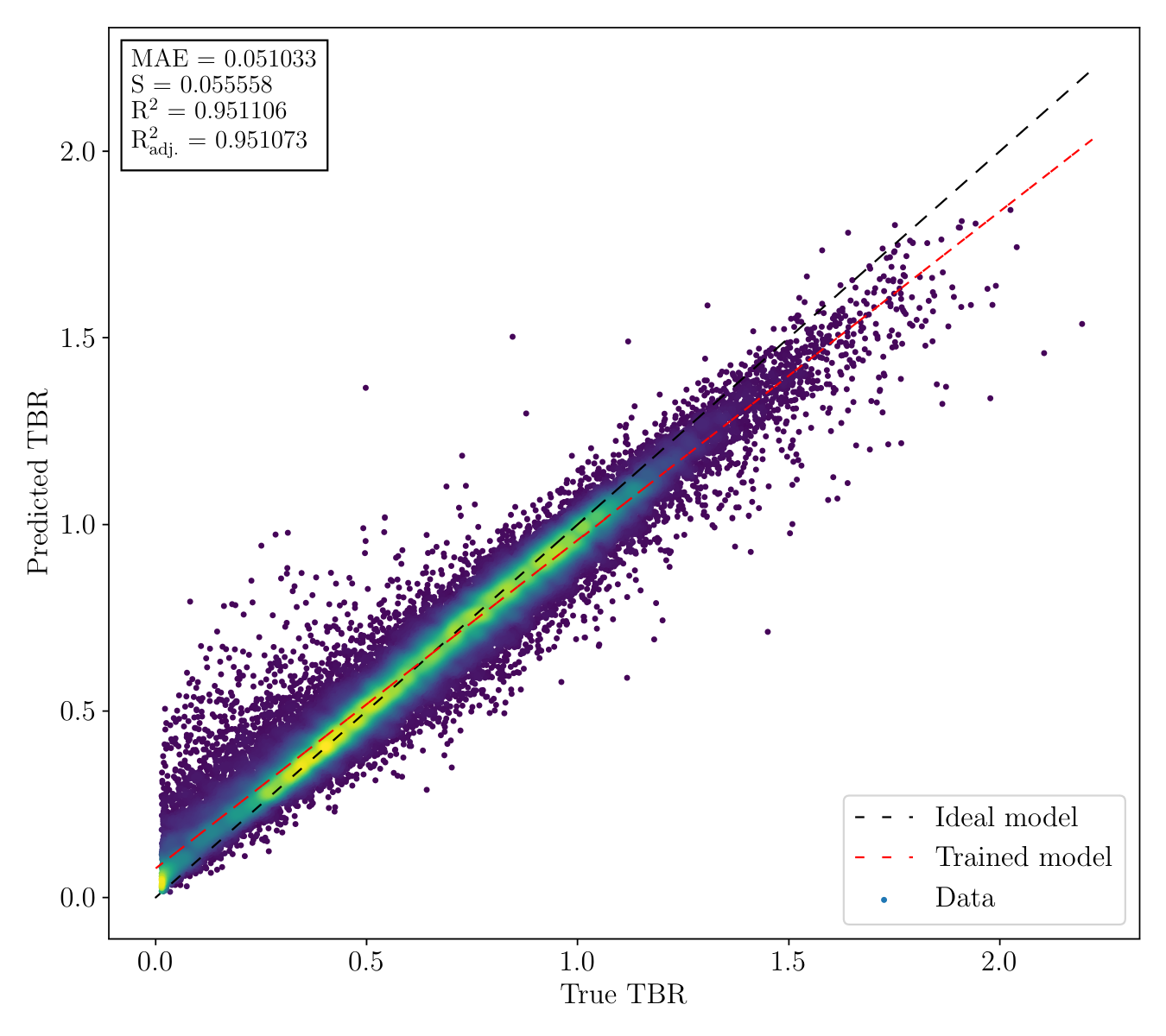}{0.051033}{0.055558}{0.951106}{0.951073}{3.82}
	\end{minipage}%
	\begin{minipage}{0.53\linewidth}
		\centering
		\surrogateplot{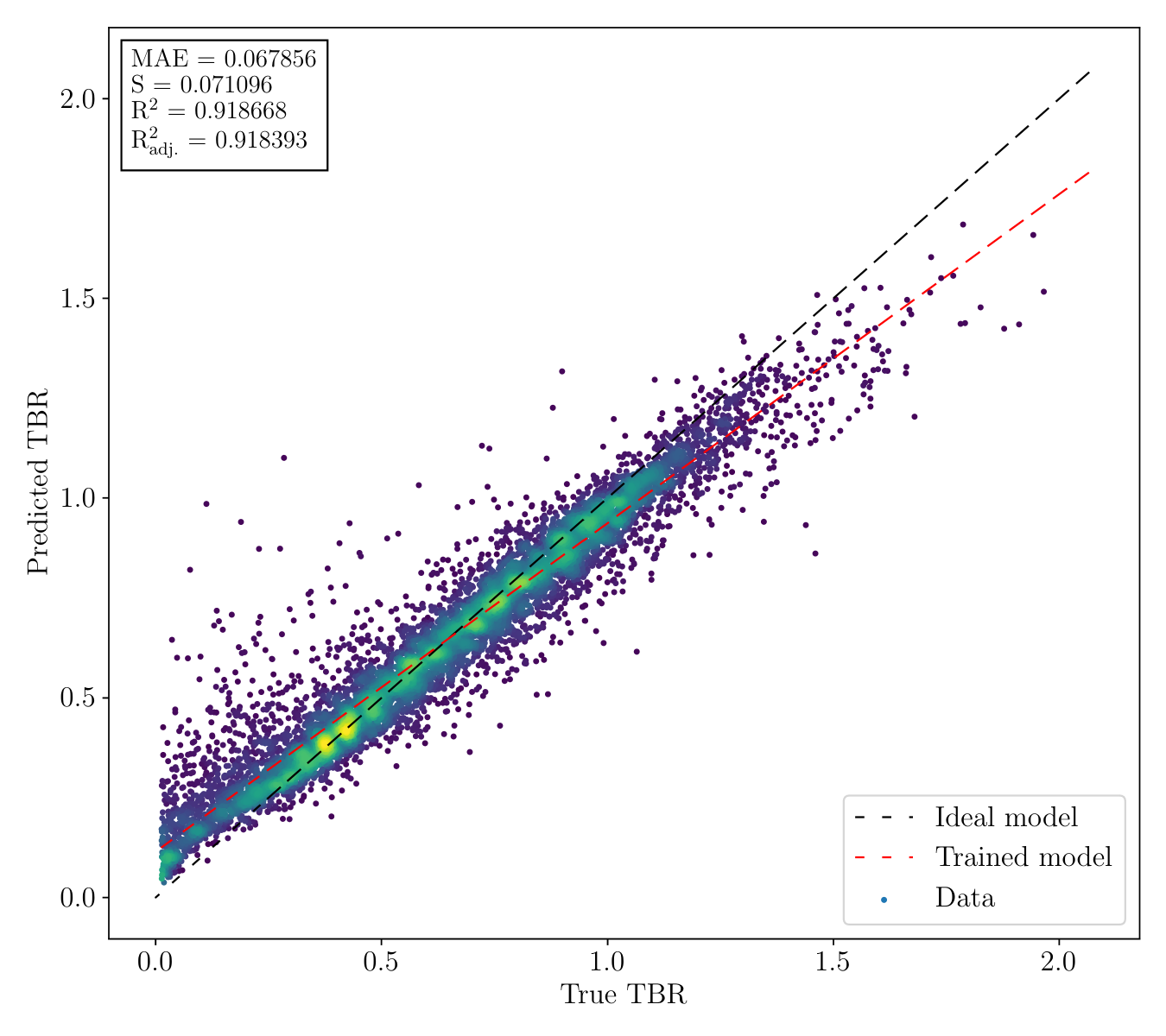}{0.067856}{0.071096}{0.918668}{0.918393}{3.82}
	\end{minipage}

	\hspace*{-0.06\linewidth}
	\begin{minipage}{0.53\linewidth}
		\centering
		\surrogateplot{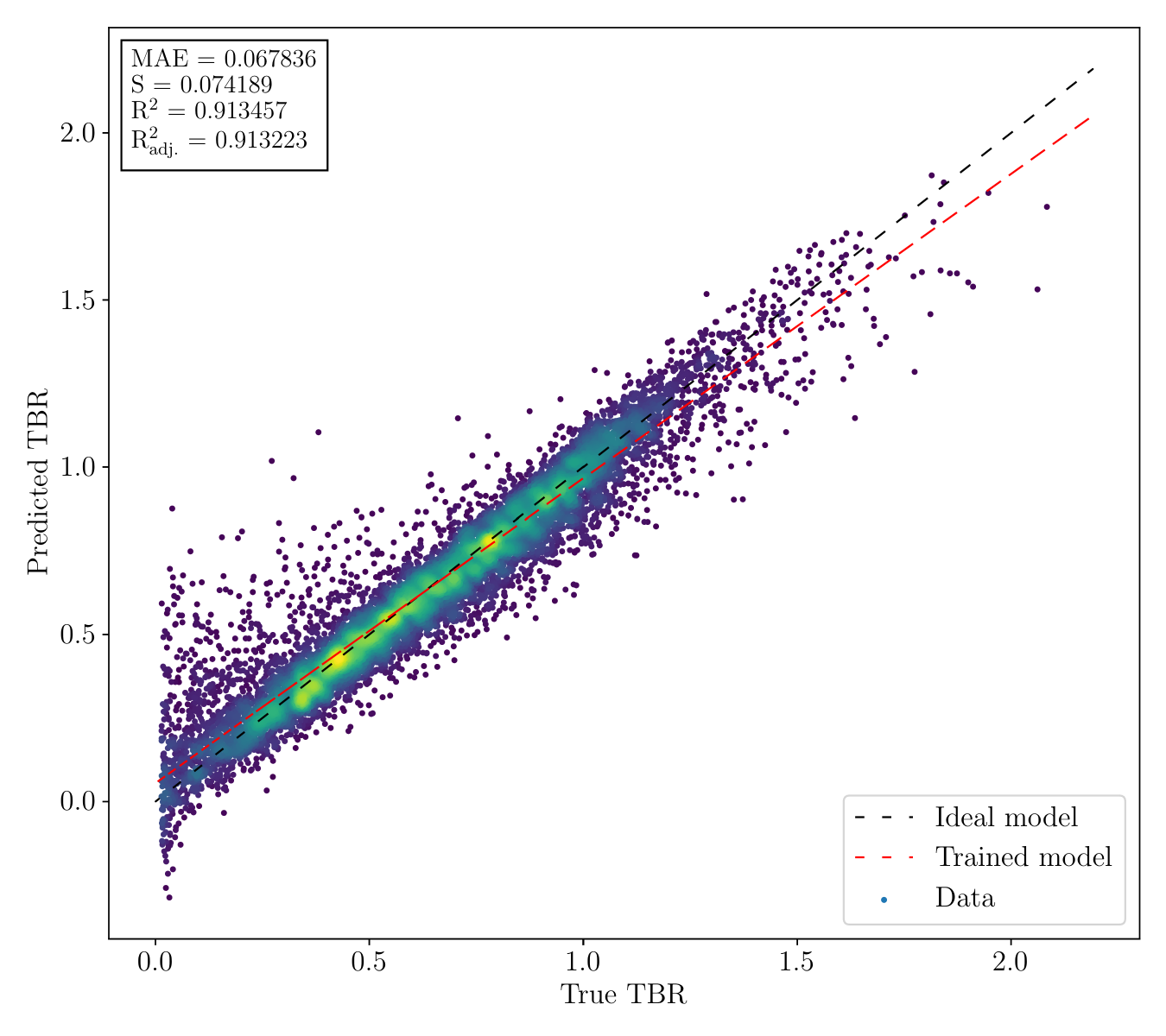}{0.067836}{0.074189}{0.913457}{0.913223}{3.82}
	\end{minipage}%
	\begin{minipage}{0.53\linewidth}
		\centering
		\surrogateplot{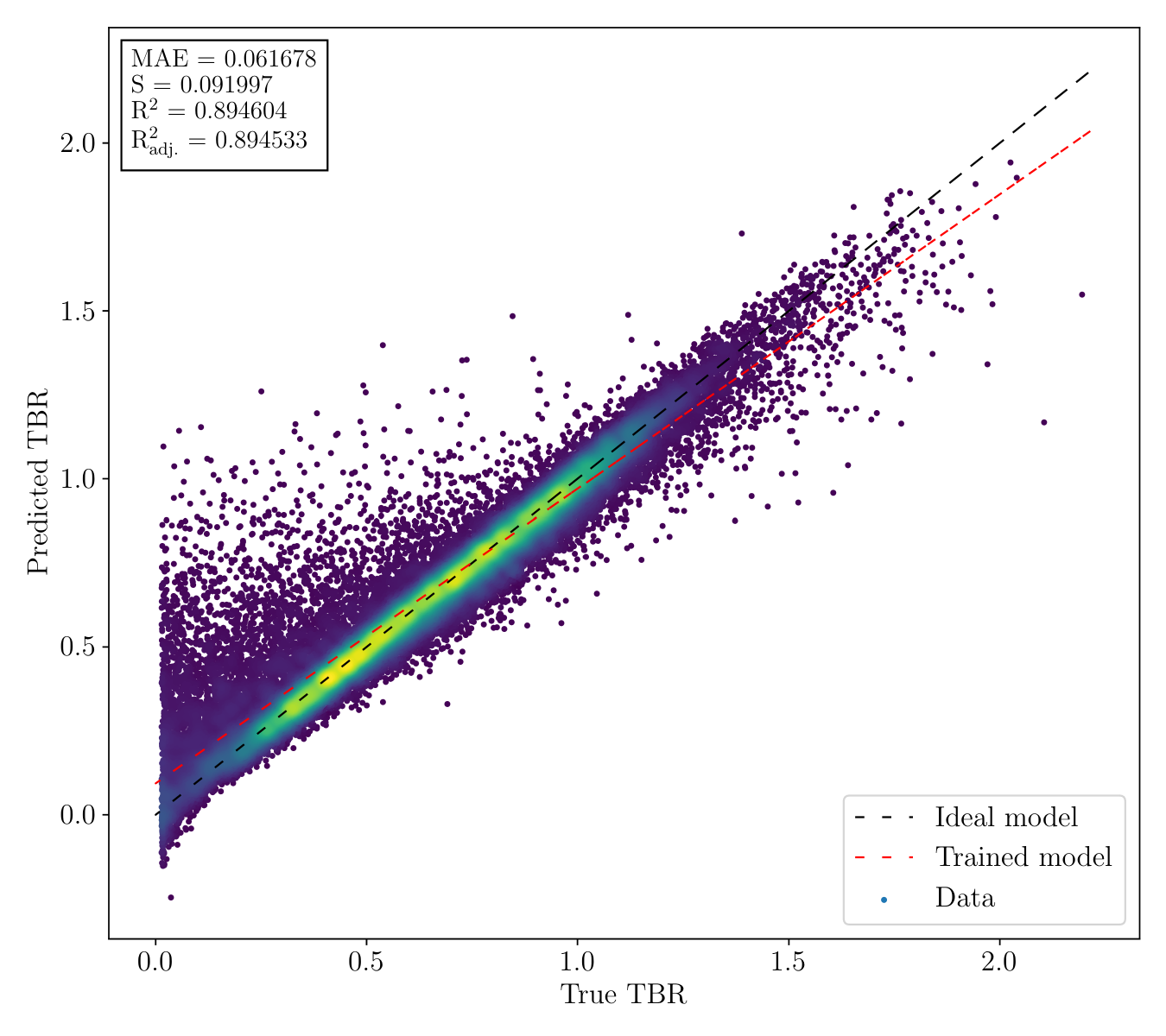}{0.061678}{0.091997}{0.894604}{0.894533}{3.82}
	\end{minipage}

	\caption{Regression performance of Models~1-8 (from left to right, top to
		bottom) in Experiment~4, viewed
		as true vs.~predicted TBR on a test set of a selected cross-validation fold. Points are colored by density.}
	\label{fig:reg-performance}
\end{figure}

{
\sisetup{round-mode=places,round-precision=3,detect-weight=true,detect-family=true}
In Experiment~4, we aim to create models that yield:
(a)~the best regression performance regardless
of other features, (b)~acceptable performance with the shortest mean
prediction time, or (c)~acceptable performance with the smallest training set.
To this end, we trained 8~surrogates that are presented in~\Fref{fig:reg-performance}
and~\Tref{tbl:exp4-detailed-results}. We compared these
surrogates with the baseline represented by Paramak per-sample evaluation time $\overline{t}_{\text{eval.}}=\num{7.777049573054314} \pm
\num{2.8103592103930337} \text{ s}$, which was measured earlier on a set of~\num{500000} samples.
}

Having selected ANNs, GBTs, ERTs, RBFs and SVMs based on the results of
Experiments~2 \&~3, we utilized the best-performing hyperparameters.
In pursuit of goal~(a), the best approximator (Model~1,
ANN) achieved~$R^2=\num{0.998}$ and mean prediction
time~$\overline{t}_{\text{pred.}}=\SI{1.124}{\micro\second}$. These correspond
to a standard error~$S=\num{0.013}$ and a relative speedup~$\omega=\num{6.92} \times {10^6}$
with respect to Paramak. Satisfying
goal~(b), the fastest model (Model~2, ANN) achieved~$R^2=\num{0.985}$,
$\overline{t}_{\text{pred.}}=\SI{0.898}{\micro\second}$, $S=\num{0.033}$
and~$\omega=\num{8.66} \times {10^6}$.
While these surrogates
were trained on the entire available set of~\num{500000} datapoints, to satisfy
goal~(c) we also trained a more simplified model (Model~4, GBT)
that achieved~$R^2=\num{0.913}$,
$\overline{t}_{\text{pred.}}=\SI{6.125}{\micro\second}$, $S=\num{0.072}$ and $\omega=\num{1.27} \times {10^6}$
with a set of size only~\num{10000}.

\begin{table*}
	\centering
	\sisetup{round-mode=places,round-precision=3,detect-weight=true,detect-family=true}
	\caption{\label{tbl:exp4-detailed-results}Results of Experiment~4. Here,
		means and standard deviations are reported over 5~cross-validation folds,
		$|\mathcal{T}|$~denotes cross-validation set size ($\times 10^3$)
		and $\omega$ is a relative speedup with respect to
		$\overline{t}_{\text{eval.}}=\num{7.777049573054314} \pm
		\num{2.8103592103930337} \text{ s}$
		measured in Paramak over~\num[round-precision=0]{500000} samples.
	The best-performing method(s) under each metric are highlighted in bold.}
	\setlength\tabcolsep{4pt}
	\renewcommand{\arraystretch}{0.95}
	\begin{indented}
	\item[]
		\scriptsize
		\begin{tabular}{lrrrrrrrr}
		\toprule
		{} & {} & \multicolumn{4}{c}{Regression performance} &
		\multicolumn{3}{c}{Computational complexity}\\
		\cmidrule(lr){3-6}
		\cmidrule(lr){7-9}
		Model & $|\mathcal{T}|$ & MAE [TBR] & $S$ [TBR] & $R^2$ [rel.] & $R^2_{\text{adj.}}$ [rel.]
						& $\overline{t}_{\text{trn.}}$ [\si{\milli\second}] &
		$\overline{t}_{\text{pred.}}$ [\si{\milli\second}] & $\omega$ [rel.]\\
		\midrule
		
		1 (ANN)
						& $\num[round-precision=0]{500.0}$
						& {\bfseries $\num{0.008777} \pm \num{0.000269}$}
						& {\bfseries $\num{0.012512} \pm \num{0.000535}$}
						& {\bfseries $\num{0.997995} \pm \num{0.000150}$}
						& {\bfseries $\num{0.997995} \pm \num{0.000150}$}
						& $\num{3.658670} \pm \num{0.035377}$
						& {\bfseries $\num{0.001124} \pm \num{0.000062}$}
						& $\num{6916416} \times$
\\

		2 (ANN)
						& $\num[round-precision=0]{500.0}$
						& $\num{0.025271} \pm \num{0.000719}$
						& $\num{0.033191} \pm \num{0.001331}$
						& $\num{0.985065} \pm \num{0.001069}$
						& $\num{0.985061} \pm \num{0.001069}$
						& $\num{2.989270} \pm \num{0.026018}$
						& {\bfseries $\num{0.000898} \pm \num{0.000037}$}
						& {\bfseries $\num{8659251} \times$}
\\

		3 (GBT)
						& $\num[round-precision=0]{200.0}$
						& $\num{0.058242} \pm \num{0.000528}$
						& $\num{0.059233} \pm \num{0.000337}$
						& $\num{0.941086} \pm \num{0.000844}$
						& $\num{0.941046} \pm \num{0.000845}$
						& $\num{2.220903} \pm \num{0.010040}$
						& $\num{0.006647} \pm \num{0.000218}$
						& $\num{1169933} \times$
\\

		4 (GBT)
						& $\num[round-precision=0]{10.0}$
						& $\num{0.070804} \pm \num{0.001843}$
						& $\num{0.071597} \pm \num{0.003491}$
						& $\num{0.913014} \pm \num{0.006027}$
						& $\num{0.911823} \pm \num{0.006110}$
						& {\bfseries $\num{1.621323} \pm \num{0.007535}$}
						& $\num{0.006125} \pm \num{0.000291}$
						& $\num{1269777} \times$
\\

		5 (ERT)
						& $\num[round-precision=0]{200.0}$
						& $\num{0.051286} \pm \num{0.000288}$
						& $\num{0.056296} \pm \num{0.000486}$
						& $\num{0.950486} \pm \num{0.000738}$
						& $\num{0.950453} \pm \num{0.000739}$
						& $\num{2.634038} \pm \num{0.009780}$
						& $\num{0.214195} \pm \num{0.003631}$
						& $\num{36308} \times$
\\

		6 (ERT)
						& $\num[round-precision=0]{40.0}$
						& $\num{0.067868} \pm \num{0.000302}$
						& $\num{0.071722} \pm \num{0.000461}$
						& $\num{0.917489} \pm \num{0.001005}$
						& $\num{0.917210} \pm \num{0.001009}$
						& $\num{2.368460} \pm \num{0.005461}$
						& $\num{0.187990} \pm \num{0.008412}$
						& $\num{41370} \times$
\\

		7 (RBF)
						& $\num[round-precision=0]{50.0}$
						& $\num{0.068405} \pm \num{0.000813}$
						& $\num{0.076889} \pm \num{0.001908}$
						& $\num{0.909963} \pm \num{0.003076}$
						& $\num{0.909719} \pm \num{0.003084}$
						& $\num{3.452536} \pm \num{0.018824}$
						& $\num{1.512068} \pm \num{0.016163}$
						& $\num{5143} \times$
\\

		8 (SVM)
						& $\num[round-precision=0]{200.0}$
						& $\num{0.062351} \pm \num{0.000493}$
						& $\num{0.094484} \pm \num{0.001577}$
						& $\num{0.890579} \pm \num{0.002923}$
						& $\num{0.890505} \pm \num{0.002925}$
						& $\num{33.346811} \pm \num{0.381933}$
						& $\num{2.415167} \pm \num{0.010751}$
						& $\num{3220} \times$
 \\
		\bottomrule
		\end{tabular}
	\end{indented}
\end{table*}

Overall we found that due to their superior performance, boosted tree-based
approaches seem to be advantageous for fast surrogate modelling on relatively small training
sets (up to the order of~$10^4$). Conversely, while neural networks perform
poorly in such a setting, they dominate on larger training sets (at least of the
order of~$10^5$) both in terms of regression performance and mean prediction time.

\subsection{Adaptive Approach}\label{sec:adaptiveres}

In order to test our QASS prototype, several functional toy theories for TBR were developed as alternatives to the expensive MC model. QASS performance was verified by training an ANN on
these theories for varied quantities of initial, incremental, and MCMC
candidate samples. By far the most useful
of these was the following sinusoidal theory, as ANNs trained on this model demonstrated similar performance to those on the expensive
MC model:

\begin{equation}
	\text{TBR} = |C|^{-1}\sum_{i \in C} \left[1 + \sin(2\pi n (x_i - 1/2)) \right]
\end{equation}

where $C$ denotes the continuous parameter space, and $n$ is an adjustable wavenumber parameter.

An increase in initial samples with increment held constant had a strong impact
on final surrogate precision, an early confirmation of basic functionality. An
increase in MCMC candidate samples was seen to have a positive but very weak
effect on final surrogate precision, suggesting that the runtime of MCMC on each
iteration could be limited for increased efficiency. We also found that an optimum increment exists and is monotonic with initial sample quantity, above or below which models showed slower improvement on both the training and evaluation sets, and a larger minimum error on the
evaluation set. This performance distinction will be far more
significant for an expensive model such as Paramak, where the number of sample
evaluations is the primary computational bottleneck.

\begin{figure}
	\centering
	\hspace*{-0.5em}\includegraphics[width=1.1\linewidth]{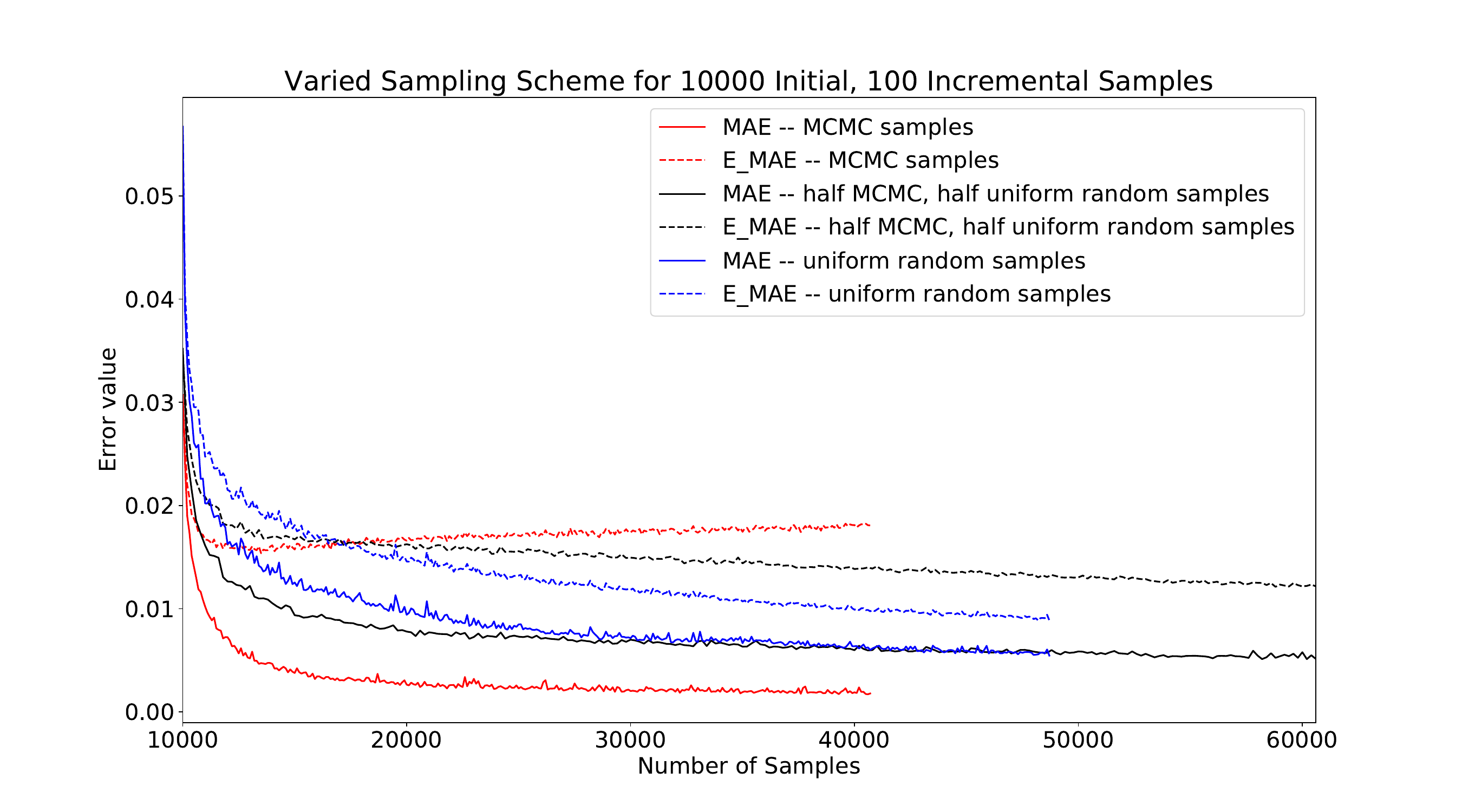}
	\caption{\label{fig:qasssampling}Absolute training error for QASS, baseline scheme, and mixed scheme.}
\end{figure}

A plateau effect in surrogate error on the evaluation set was universal to all configurations, and initially suspected to be a residual
effect of retraining the same ANN instance without adjustment to data
normalisation. A ``Goldilocks scheme'' for checking normalisation drift was
implemented and tested, but did not affect QASS performance. Schemes in which
the ANN is periodically retrained were also discarded, as the retention of
network weights from one iteration to the next was demonstrated to greatly
benefit QASS efficiency. Further insight came from direct comparison between
QASS and a baseline scheme with uniformly random incremental samples, shown
in~\Fref{fig:qasssampling}.

\begin{figure}
	\centering
	\hspace*{-0.5em}\includegraphics[width=1.1\linewidth]{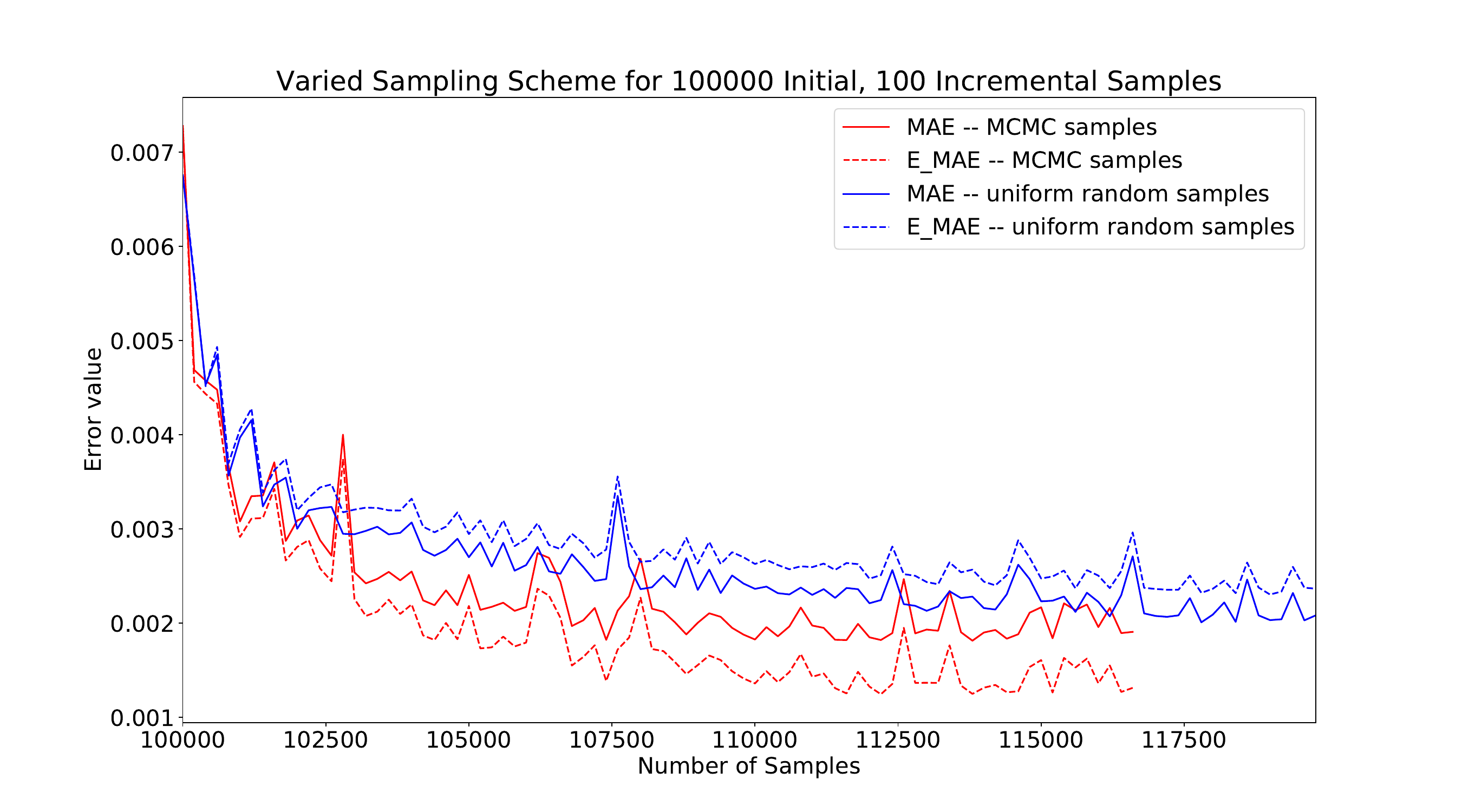}
	\caption{\label{fig:qasssampling100k}Absolute training error for QASS and
	baseline scheme, with~\num{100000} initial samples.}
\end{figure}

Such tests revealed that while QASS has unmatched performance on its own
adaptively-sampled training set, it is outperformed by the baseline scheme on
uniformly-random evaluation sets. We inferred that while QASS excels in
learning the most strongly peaked regions of the TBR theory, this comes at the
expense of precision in broader, smoother regions where uniformly random
sampling suffices. Therefore a mixed scheme was implemented, with half MCMC
samples and half uniformly-random samples incremented on each iteration, which
is also shown in~\Fref{fig:qasssampling}. An increase in initial sample size was also observed to improve precision in these smooth regions of the toy theory, as
the initial samples were uniformly-random. As shown
in~\Fref{fig:qasssampling100k}, with~\num{100000} initial samples it was
possible to obtain a ${\sim}40\%$ decrease in error as compared to the baseline
scheme, from 0.0025 to 0.0015 mean averaged error. Comparing at the point of
termination for QASS, this corresponds to a ${\sim}6\%$ decrease in the number
of total samples needed to train a surrogate model while achieving the same error.

\section{Conclusion}
\label{sec:conclusion}
We employed a broad spectrum of data
analysis and machine learning techniques to develop a library of fast and high-quality
surrogate models for the expensive Paramak TBR model. After reviewing
9~surrogate model families, examining their behaviour on a constrained and
unrestricted feature space, and studying their scaling properties, we retrained
the best-performing instances to produce properties desirable for
practical use. The fastest surrogate, an artificial neural network trained
on~\num{500000} datapoints, attained an~$R^2=\num{0.985}$ with mean prediction time
of~$\SI{0.898}{\micro\second}$, representing a relative
speedup of $8\cdot 10^6$ with respect to Paramak. Furthermore, we demonstrated the possibility of achieving comparable results using only a
training set of size~\num{10000}.

We additionally developed a novel adaptive
sampling algorithm, QASS, capable of interfacing with any of the surrogate models presented in this work.
Preliminary testing on a toy theory, qualitatively comparable to
Paramak, demonstrated the effectiveness of QASS and key behavioral trends. With~\num{100000} initial samples
and 100 incremental samples per iteration, QASS achieved a ${\sim}40\%$ decrease
in surrogate error compared to a baseline random sampling scheme. Further optimisation over the hyperparameter space has strong potential to increase this performance by further reduction of necessary expensive samples, in particular by decreasing the required quantity of initial samples. This will allow for future deployment of QASS on top of any of our most effective identified TBR surrogate models.

Relevant source code, model instances and datasets are freely
available online as well as a more detailed technical
report~\cite{github,finalreport}.


\section{Acknowledgements}
\label{sec:acknowledgements}
PM \& GVG were supported by the STFC UCL Centre for Doctoral Training in Data Intensive
Science (grant no. ST/P006736/1). GVG was funded by the UCL Graduate Research and Overseas Research Scholarships.

This project was supported by the EU Horizon 2020 research \& innovation
programme [grant No 758892, ExoAI]. N.~Nikolaou acknowledges the support of the NVIDIA Corporation’s GPU grant.

This work has been carried out within the framework of the EUROfusion consortium and has received funding from the Euratom research and training programme 2014-2018 and 2019-2020 under grant agreement No 633053. The views and opinions expressed herein do not necessarily reflect those of the European Commission.

This work has been partly funded by the Institutional support for the
development of a research organization (DKRVO, Czech Republic).

This work has also been part-funded by the RCUK Energy Programme (grant number EP/I501045/1).

\section{References}
\label{sec:references}
\bibliography{ucl_tbr_paper}
\bibliographystyle{iopart-num}

\end{document}